\documentclass[useAMS,usenatbib,usegraphicx]{mn2e}
\usepackage{aas_macros,times}

\title[Stripe 82 SFRDs]{The local star-formation rate density: assessing calibrations using \oii, \ha~and UV luminosities}
\author[D. G. Gilbank et al.]{David G.~Gilbank$^1$\thanks{Email: dgilbank@astro.uwaterloo.ca}, Ivan K. ~Baldry$^2$, Michael L.~Balogh$^1$,
Karl~Glazebrook$^3$ 
\newauthor
and Richard G.~Bower$^4$
\newauthor\\
  $^1$Department of Physics and Astronomy, University of Waterloo, Waterloo, Ontario, Canada N2L 3G1\\
  $^2$Astrophysics Research Institute, Liverpool John Moores University, Twelve Quays House, Egerton Wharf, Birkenhead CH41 1LD, UK\\
  $^3$Centre for Astrophysics and Supercomputing, Swinburne University of Technology, P.O. Box 218, Hawthorn, VIC 3122, Australia\\
  $^4$Institute for Computational Cosmology, Department of Physics, University of Durham, South Road, Durham, DH1 3LE, UK\\
}

\pagerange{\pageref{firstpage}--\pageref{lastpage}} \pubyear{2007}

\def\LaTeX{L\kern-.36em\raise.3ex\hbox{a}\kern-.15em
    T\kern-.1667em\lower.7ex\hbox{E}\kern-.125emX}
\def\gsim{\mathrel{\raise0.35ex\hbox{$\scriptstyle >$}\kern-0.6em
\lower0.40ex\hbox{{$\scriptstyle \sim$}}}}
\def\lsim{\mathrel{\raise0.35ex\hbox{$\scriptstyle <$}\kern-0.6em
\lower0.40ex\hbox{{$\scriptstyle \sim$}}}}

\def\ha{{\rm H$\alpha$}}
\def\hb{{\rm H$\beta$}}
\def\oii{[{\sc O\,II}]}
\def\oiii{[{\sc OIII}]}

\def\b04{B04}

\def\ms{$\log(M_*/M_\odot)$}
\def\sfrtot{$SFR_{\rm Tot}$}
\def\sfrfib{$SFR_{\rm fibre}$}
\def\msun{M$_\odot$}

\def\lsim{\mathrel{\hbox{\rlap{\hbox{\lower4pt\hbox{$\sim$}}}\hbox{$<$}}}}
\def\gsim{\mathrel{\hbox{\rlap{\hbox{\lower4pt\hbox{$\sim$}}}\hbox{$>$}}}}

\begin{document}

\label{firstpage}

\maketitle

\begin{abstract}
We explore the use of simple star-formation rate (SFR) indicators (such as may be used in high-redshift galaxy surveys) in the local Universe using \oii, \ha, and $u$-band luminosities from the deeper 275 deg$^2$ Stripe 82 subsample of the {\it Sloan Digital Sky Survey} (SDSS) coupled with UV data from the {\it Galaxy Evolution EXplorer satellite} ({\it GALEX}).  We examine the consistency of such methods using the star-formation rate density (SFRD) as a function of stellar mass in this local volume, and quantify the accuracy of corrections for dust and metallicity on the various indicators.  Rest-frame $u$-band promises to be a particularly good SFR estimator for high redshift studies since it does not require a particularly large or sensitive extinction correction, yet yields results broadly consistent with more observationally expensive methods.  We suggest that the \oii-derived SFR, commonly used at higher redshifts (z$\sim$1), can be used to reliably estimate SFRs for ensembles of galaxies, but for high mass galaxies (\ms$\gsim$10), a larger correction than is typically used is required to compensate for the effects of metallicity dependence and dust  extinction. We provide a new empirical mass-dependent correction for the \oii-SFR. 
{\bf Note:  This astro-ph version has been updated to correct the typographical errors in equations 2, 7, and 9 of the originally published paper, as described in the MNRAS Erratum.}
\end{abstract}


\section{Introduction}
The most useful observables in constructing a picture of galaxy formation are likely to be: a galaxy's star formation rate (SFR), its stellar mass, and the epoch at which it is observed.  Stellar masses are now routinely measured out to z$\sim$4 (e.g., \citealt{Marchesini:2009fx}) by fitting spectral energy distributions (SEDs) of stellar population synthesis models to multicolour broadband photometry.  Although the uncertainties in stellar mass estimates are dominated by evolutionary uncertainties within the models \citep{Marchesini:2009fx}, most workers use the same models and similar passbands for the photometry (or the differences between models are at least well understood) and thus the comparison between stellar masses used in different works is relatively straightforward. The situation is somewhat less uniform for SFR measurements.  Indeed, most of the systematic errors affecting SFR measurements are likely to be functions of stellar mass \citep{Brinchmann:2004ct,Kewley:2004vs,Cowie:2008ob,Pannella:2009uj}.

Over the last decade, measuring the star formation rate density (SFRD) of large samples of galaxies out to z$\sim$6 has become routine, and dozens of such measurements now exist \citep{Lilly:1996la,Madau:1996ao,1999ApJ...519....1S,Hopkins:2006bv,Reddy:2009wc}.  In this time of  `survey science', statistical samples are commonplace and the largest remaining uncertainties in such studies are systematic in nature, for example, the choice of SFR indicator. Common proxies for SFR include the luminosity of emission lines such as \ha, or the UV continuum, which are sensitive to the ionising flux from young stars; or mid- or far-infrared radiation (MIR, FIR) which measure the fraction of this flux absorbed and re-radiated by dust. 

Each SFR indicator possesses its own strengths and disadvantages.  The luminosity of the \ha~recombination line is relatively simply and directly coupled to the incident number of Lyman continuum photons produced by young stars, and hence is proportional to the SFR \citep{Kennicutt:1998pa}, although there is some dependence on the metallicity of the gas \citep{Charlot:2001gj}.  Provided the line can be adequately corrected for stellar absorption, the largest uncertainty is the correction for the presence of dust.  At higher redshifts, $z\gsim0.4$, the \ha~line passes out of the optical window and the most commonly-used emission line in the range $0.4 \lsim z \lsim 1.5$ has been \oii$\lambda 3727$.  \oii~luminosity is more strongly affected by dust than \ha.  Moreover, this collisionally-excited line is very sensitive to metallicity.   Since the emission lines necessary to accurately measure dust extinction and metallicity are not available when \oii~is used (otherwise the \ha~line would be used in preference to \oii), alternate approaches to empirically correct the \oii-SFR for these dependencies have been adopted \citep[e.g.,][]{Moustakas:2006wp,Weiner:2007tf}.  Another observational limitation of indicators based on spectroscopy, particularly for low redshift, is the need for an aperture correction.  The spectroscopic aperture may only sample the innermost regions of the galaxy and thus an extrapolation is necessary to sample the total light from the galaxy.  This may be further complicated by population gradients across the galaxy and the fixed angular size aperture sampling a different fraction of  a given galaxy at different redshifts.  A comprehensive examination of aperture effects within SDSS data is given in \citet{Brinchmann:2004ct}.

The rest-frame UV luminosity is straightforward to measure in high redshift surveys.  Indeed, it is normally produced as a by-product of the survey data, since all that is required is a relatively blue optical band (depending on the redshift of the source).   The UV luminosity provides a measurement of the  continuum radiation from young stars, again after it has been extinguished by dust.  Although sensitive to dust, the UV is relatively insensitive to metallicity (e.g., \citealt{Glazebrook:1999kl}).  Additionally, a correction must be made for the contribution to the UV luminosity from more evolved stellar populations.  This correction becomes larger towards lower redshifts, as the Universe ages and so do the typical ages of the galaxies considered.  The extinction due to dust may be estimated from the slope of the UV continuum (which is easily measured from two or more filter photometry).  This method was originally applied, with considerable uncertainty, only to starburst galaxies \citep{Meurer:1999qd}, but recent work has found comparable relations for normal galaxies (e.g., \citealt{Cortese:2006rm}).  The effect of dust on these indicators can be large and varies considerably from galaxy-to-galaxy.  In the local Universe, the dust extinction in \ha~can be 0-2 mag and in the UV 0-4 mag \citep{Kennicutt:1998pa,Calzetti:2000sp}.  This scatter may be partly driven by galaxy inclination and dust geometry effects, meaning that even for galaxies with the same dust content, a different effective attenuation may be observed.

Furthermore, the different indicators are sensitive to the presence of young stars for different lengths of time: the ionising radiation responsible for the \ha~line measures SFRs on timescales of $\sim$10 Myr, whereas UV-bright stars which are considered in the SFR derived from UV luminosity have lifetimes $\sim$100 Myr.  Similarly, the sensitivity of the ultraviolet continuum and emission line luminosities to stars of different masses ($\gsim5$ and $\gsim10$ M$_\odot$ respectively) means that the derived SFRs are sensitive to the IMF. This also means that if the assumption of a universal IMF is incorrect, this will lead to differences between these SFR indicators (e.g., \citealt{Meurer:2009la}).

Thus, the SFR is difficult to estimate accurately for an individual galaxy.  One solution is to measure the SFRs for large ensembles of galaxies and compare their average properties in order to smooth over some of these effects.  For example, comparing globally-averaged SFR quantities estimated from different indicators with different sensitivities to dust, metallicity, etc. should show the same results, on average,  if the corrections for these effects are broadly consistent. Now, the potential systematic error of most concern is the timescale to which the different indicators are sensitive. However, provided the timescales of all the indicators used are short compared with the timescale over which the cosmic average SFR is changing, this difference can be negated.

Since most of the variables affecting the estimate of SFR correlate with mass, e.g., metallicity via the mass--metallicity relation \citep{Tremonti:2004ik}, dust \citep{Brinchmann:2004ct}; an obvious first step is to determine mass-dependent corrections to the SFR indicators.  Furthermore, there has been great activity recently in measuring how SFR (or some related quantity such as SFRD or SFR/stellar mass) varies as as a function of stellar mass, and how this evolves with redshift.  Most of these results have been consistent with a picture in which the termination of star-formation progresses from higher mass to lower mass galaxies as the Universe ages, so-called {\it cosmic downsizing} \citep{Cowie:1996xw}.  Obviously any residual systematic errors as a function of mass on the SFR indicators used could pose a serious problem when studying such relatively subtle mass-dependent effects.  

To date, most high-redshift studies have used locally-calibrated tracers of SFR, under the assumption that the calibration is still valid at higher redshift.  Most local SFR calibrations use a simple conversion between luminosity of some SFR proxy and SFR and assume that this is valid for all masses.  However, as we will show in this paper, a better approach is to allow for a mass-dependent conversion factor which accounts for differences such as the average dust extinction as a function of galactic mass.  Dust obscuration at higher redshift is an open question.  For example, in two recent z$\sim$2 studies, \citet{Pannella:2009uj} and \citet{Reddy:2009wc} both find evidence for mass-dependent extinction, but the former find an extremely large dust correction is necessary (much larger than that measured locally, \S~\ref{sec:othemp}), whereas the latter find that the amount of dust extinction required is a {\it decreasing} function of redshift.  Part of this difference may be due to sample selection in that the \citet{Reddy:2009wc} sample selects Lyman Break Galaxies, which may favour systems with lower UV extinction, but this still goes to show that the evolution of dust properties is not well constrained. Other works (e.g., \citealt{Martin:2007xe,Iglesias-Paramo:2007db}) at z$\sim$1 find similar evidence for mass-dependent extinction using the ratio of IR to FUV luminosity to estimate extinction.

Regardless, a sensible null hypothesis would be to assume that the higher redshift relations (such as the mass dependence of extinction) follow the local relations, and derive SFR calibrations and corrections based on local data.  Using a local mass-dependent correction at higher redshift is clearly preferable to ignoring the mass dependence of SFR indicators altogether.  As a first step in understanding how the various SFR indicators commonly used at higher redshift relate to one another, we have undertaken a comparison between these indicators in the local Universe.  The Sloan Digital Sky Survey (SDSS, \citealt{york}) provides high quality optical photometry and spectra for a well-understood sample of galaxies.  In addition, several groups have already performed detailed measurements of SFRs using state-of-the-art techniques (fitting spectrophotometric data to the latest stellar evolutionary synthesis and photoionisation models) which may be used as a reference against which simpler methods, such as those necessarily employed at higher redshift, may be compared.  
 
The main aim of this paper is to examine the consistency between different star-formation indicators by exploring the SFRD estimates obtained with the different tracers and to look for possible systematic differences as a function of mass. We examine these in the local Universe, but end with a brief exploration of how these may affect higher redshift surveys. This paper is designed to serve as a local reference, both for understanding how the various estimators compare with each other and to examine evolution when used in conjunction with higher redshift surveys.  We proceed in the following way. In \S2 we describe the SDSS optical imaging and spectroscopy and {\it GALEX} UV photometry. \S3 describes the SFR indicators examined. We use the \ha~and \oii~emission lines and UV photometry from the SDSS $u$-band and {\it GALEX} FUV. For each of these indicators, we explore a range of different assumptions for the dust correction, e.g., a nominal constant extinction, extinction estimated from optical emission lines (or, for example, the UV slope). We present results for our `best estimator' for each indicator throughout the paper, but also show how these estimates change under the various different assumptions commonly made. Many of the prescriptions we use are well known in the literature (particularly the UV), but we will show that current \oii~indicators are far from optimal and first we will need to correct this.  In \S4 we present our new empirical (mass-dependent) correction to \oii; \S5 examines the emission line luminosity functions and \S6 presents the the SFRD results, convergence tests for the limiting SFR necessary to accurately estimate the SFRD, and also briefly considers the application of these to a higher redshift survey.  A more detailed z$\sim$1 analysis drawing on these local results will be presented in \citet{Gilbank:2010hc}. In \S7 we summarise our conclusions.  

All magnitudes are quoted on the AB system unless otherwise stated, and we assume a cosmology $(h,\Omega_M, \Omega_\lambda)=(0.7,0.3,0.7)$. Throughout, we convert all quantities to those using a \citet{Kroupa:2001ea} universal initial mass function (IMF), except for \S\ref{sec:highz}.

\section{Data and method}
\label{sec:data}

The SDSS is a project \citep{york} that has imaged $10^4$ deg$^2$ and obtained spectra of $10^6$ objects \citep{sdssdr7}. The imaging was obtained using five broadband filters by drift scanning with a mosaic CCD camera on a 2.5-m telescope \citep{Gunn:1998ew}.  Various samples were selected for spectroscopic targeting using a 640-fibre fed spectrograph on the same telescope. The samples include an $r<17.8$ main galaxy sample (MGS; \citealt{Strauss:2002ys}) and a colour-selected $r<19.5$ luminous red galaxy (LRG: \citealt{Eisenstein:2001xg}) sample.

The imaging survey covered $\sim8000$ deg$^2$ of the NGP. Only three stripes (scans each 2.5-deg wide) were observed in the SGP as part of the `legacy survey'. One of these, known as Stripe~82, here defined as RA from $-50.8$ to 58.6 degrees and DEC between $-1.26$ and 1.26 degrees covering an area of 275 deg$^2$, has been imaged multiple times \citep{sdssdr5}.  In this paper, we only use data from the deeper Stripe 82 region in order to probe to the lowest possible stellar masses.  For our imaging catalogue, we use the coadded-catalogue technique and data set described in \citet{Baldry:2005up}.  At the time, there were between 6 and 18 repeat images depending on the sky position. The data set uses average fluxes determined at the catalogue level, providing higher S/N magnitudes and colours. This is sufficient for our purposes because the surface brightness limit of SDSS single-scan imaging \citep{Blanton:2005gn} is probably
not missing any significant detections for galaxies with stellar masses $>10^{8.5}$\msun \citep{Baldry:2008sj}; and while there is a catalogue derived from Stripe~82 coadded imaging in DR7, the reliability of the earlier coadd at the catalogue level has been thoroughly checked.

The catalogue-level coadd is described in \S~3.1 of \citet{Baldry:2005up}.  This was for the $u$-band Galaxy Survey.  The method applied here is the same, except the requirement for fluxes with S/N
$>3$ is applied to the $g$- and $r$-band Petrosian measurements but not the $u$-band. For this paper, we select sources with $r_{psf} - r_{model}>0.2$ and $r_{Petro}<19.5$. In addition, objects were
selected to be not saturated, which is a standard flag check to remove artefacts and sources near bright stars. This produced a catalogue of 204\,093 objects.

In addition to the multiple-scan imaging, various additional spectroscopic targeting programs were implemented on Stripe~82.  In addition to the MGS and LRG selections, galaxy targeting included the
$u$-band Galaxy Survey, a low-$z$ ($<0.15$) and high-$z$ ($>0.3$) selection, and fainter LRG selection.  These are described in \S~3 of the SDSS DR4 paper \citep{sdssdr4}.  Notably the low-$z$ program also includes random sampling to $r<19.5$. From the SDSS DR7 data release, 73\,035
spectra were matched to the $r<19.5$ catalogue used in this paper. Of these, 46\,279 are for sources fainter than $r>17.8$. While these are fainter targets than the MGS, the redshift success rate is still high
and we do not make any correction for missed redshifts.

\begin{figure*}
{\centering
\includegraphics[width=140mm]{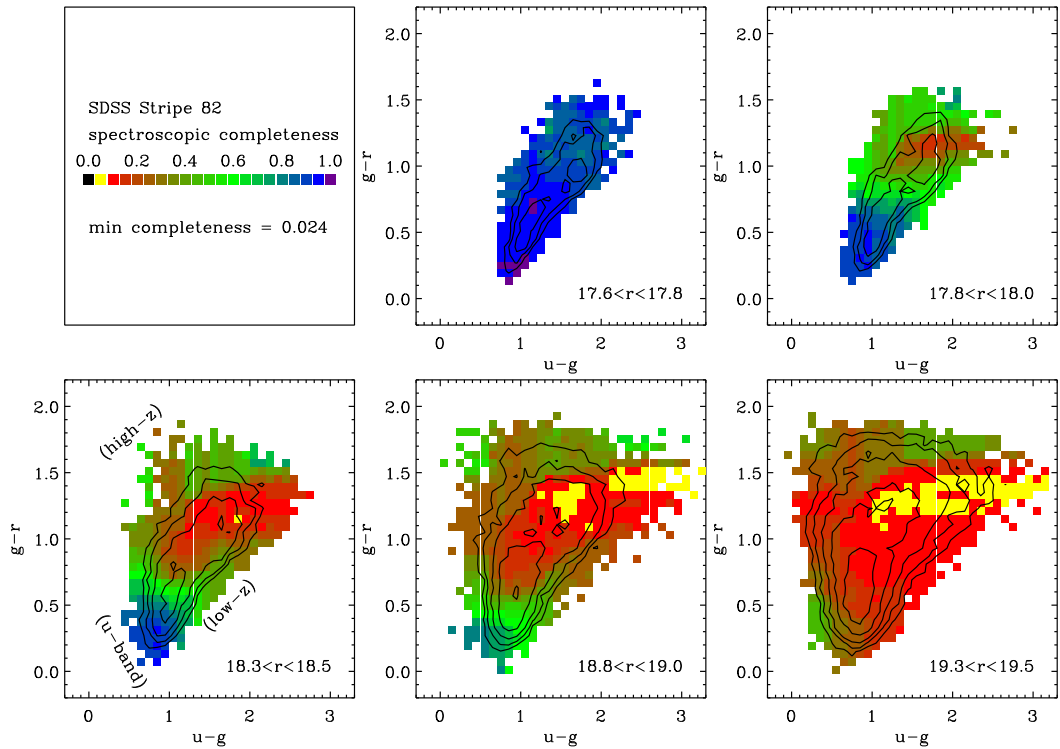}
\caption{Completeness maps: average targeting completeness as a function of
$u-g$ and $g-r$ for five magnitude ranges shown by the coloured squares. The
observed colour-colour bivariate distributions are represented by the black
contours. The completeness patterns in the lower panels are primarily the
result of the $u$-band, low-z and high-z galaxy selections (annotated in the
lower-left panel as a guide).
\label{fig:comp-maps.ps}
}
}
\end{figure*}

The targeting completeness of our sample is nearly 100\% at $r<17.8$ and drops significantly above that. To model this, given the heterogeneous selection, we divide the sample by $r$-band magnitude
and by the colours $u-g$ and $g-r$ similarly to \citet{Baldry:2005up}. Each bin contains 60--166 objects and the completeness, $c_i$, of sources within a bin is defined as the number of objects with an SDSS
spectrum divided by the number in the bin. Colour-colour completeness maps are shown in Fig.~\ref{fig:comp-maps.ps}.  Where the data are only available for spectroscopy in the DR4 release, the $c_i$ values are recomputed accordingly.

We select SDSS galaxies in the redshift range $0.032<z<0.2$, where the low redshift cut is imposed to ensure that  \oii~falls within the spectral range of the SDSS spectrograph.  The MPA group has produced value-added data products for the SDSS, however not all products are available for all generations of the survey.  Currently, emission line fluxes are available for DR7, but stellar masses and spectral classes (described below) are only available up to DR4. As such, we choose to derive the missing properties using simple scalings (which we detail below) from the existing data where necessary.  
The Stripe 82 data are mostly covered by DR4; out of the 48\,211 galaxies within our redshift range, 43\,155 have MPA-derived quantities from DR4.

\subsection{Stellar masses}
\label{sec:mass}
Stellar masses, $M_*$, for DR4 galaxies\footnote{http://www.mpa-garching.mpg.de/SDSS/DR4/} are measured as described in \citet{Kauffmann:2003ek}.  Briefly, these use the $z$-band photometry to measure the galaxy's luminosity.  A galaxy's mass-to-light ratio, $M/L$, is highly dependent on its star-formation history, SFH.  Constraints on the SFH  are obtained from the galaxy's location in the $D_n(4000)$ -- $H\delta_A$ plane, using a suite of stellar population synthesis models \citep{Bruzual:2003de} to marginalise over a range of star-formation histories.  

For the galaxies in DR7 not present in DR4, these stellar masses are not available, and so we use a luminosity and colour-based estimate of the mass-to-light ratio (see e.g., \citealt{Bell:2003iq}) fitted empirically against the \citet{Kauffmann:2003ek} masses for this small subsample of objects.  The stellar mass can be estimated to within 0.21 dex (1$\sigma$ scatter) of the  \citet{Kauffmann:2003ek} mass using 
\begin{equation}
\log(M_*/M_\odot) = 0.480(g-r)_0 -0.475 M_z -0.08
\end{equation}
where $(g-r)_0$ is the rest-frame $(g-r)$ colour\footnote{we use {\tt modelmag} to measure colours throughout} and $M_z$ is the rest frame absolute $z$-band magnitude.  Rest-frame magnitudes are estimated by $k$-correcting observer-frame magnitudes using $k$-corrections determined using the public code {\sc kcorrect v4\_1\_4} \citep{Blanton:2007if}.  Other colour combinations give slightly larger scatter.\footnote{The scatter in the colour-estimated mass is slightly smaller using magnitudes and colours transformed to z$=$0 rather than z$=$0.1 (even though the latter involves a smaller correction and should be less uncertain).  This may be in part due to the way in which the \citet{Kauffmann:2003ek} masses are calculated.  Regardless, we present the relation in terms of z$=$0 values, which should also simplify comparison with other works.}

We also estimate stellar masses for all galaxies using SED fits to the $ugriz$ photometry at the spectroscopic redshift using the technique of \citet{Glazebrook:2004zr}, such as might be done at high redshift.  Our results are negligibly affected if we substitute either the simple colour masses or the SED-fit stellar masses in place of the \citet{Kauffmann:2003ek} masses.  So, throughout this paper we use the \citet{Kauffmann:2003ek} masses, with colour masses used for the small fraction not possessing measurements of these (DR7 galaxies not present in DR4).  Similarly, if the small number of colour-based masses are neglected and we just use the DR4 masses from \citet{Kauffmann:2003ek} our results are unchanged.

\subsection{Line fluxes}
\label{sec:linefluxes}

Line fluxes are taken from the DR7 version of the MPA database\footnote{http://www.mpa-garching.mpg.de/SDSS/DR7/} and the measurement method is described in \citet{Tremonti:2004ik}.  Briefly, the spectra are compared with a suite of high-resolution stellar population synthesis models \citep{Bruzual:2003de} to fit the continuum shape, thus accounting for weak metal absorption under forbidden lines and Balmer absorption.  After accounting for absorption and removing any remaining sky residuals, Gaussian profiles are simultaneously fitted to the emission lines, requiring that all the lines in a given class (Balmer, forbidden) have the same width and velocity offset.  This method is therefore beneficial for measuring weak lines, since it uses constraints from multiple/stronger lines.  In this paper, we will primarily be interested in the lines \ha~$\lambda$6563 (hereafter \ha) and \oii~$\lambda\lambda$ 3726, 3729.  We sum the fluxes of the two components of the \oii~3727 doublet, adding their uncertainties in quadrature, and hereafter refer to this flux simply as \oii.

The MPA group derived empirical corrections to the line flux uncertainties via repeat observations\footnote{see http://www.mpa-garching.mpg.de/SDSS/DR7/raw\_data.html}.  \ha~error estimates are scaled by a factor of 2.473 and \oii~error estimates are scaled by a factor of 2.199.  These error estimates are used when selecting galaxies, as we will essentially construct \ha- and \oii-selected samples by requiring $S/N>2.5$ in the corresponding emission line flux, for each of these samples.

The SDSS is a fibre-based spectroscopic survey and thus, all spectral measurements are based on the flux through the fibre aperture.  In order to correct these quantities to something approximating total fluxes, we estimate aperture corrections by comparing an estimate of each galaxy's total magnitude (as given by the Petrosian magnitude) with the corresponding magnitude measured through the fibre.  The correction factor in magnitudes is thus given by the difference of these two measurements.  We use the fibre and Petrosian magnitudes in the $u$-band as this is likely to most accurately trace the light from star formation.  We discuss the aperture corrections in more detail in Appendix \ref{sec:apcorr}.

Ratios of line fluxes (which are thus insensitive to aperture corrections) can be used to classify the nature of the emission arising from each galaxy, the so-called BPT diagram \citep{Baldwin:1981ci}.  \citet{Kauffmann:2003af} presented a modified version of the BPT diagram to classify galaxies into star-forming, AGN and composite types.  Here we adopt the \citet{Kauffmann:2003af} scheme to provide a spectral type for each galaxy.  For the bulk of the analysis we will not separate galaxies by type, but will study the star-formation rate density of all galaxies assuming that they each follow the line luminosity--star formation rate relations discussed below.  In \S\ref{sec:class} we discuss the impact of the different classes on the sample.

\subsection{{\it GALEX} photometry}

UV data is taken from the {\it Galaxy Evolution Explorer} satellite ({\it GALEX}).  {\it GALEX} observes in two bands: the near-UV (NUV) and far-UV (FUV) at effective wavelengths of 2271\AA~and 1528\AA~respectively.  A {\it GALEX}-SDSS matched catalogue is constructed in a manner similar to that described in \citet{Budavari:2009tg}.  Briefly, the {\it GALEX} database{\footnote{see http://galex.stsci.edu/casjobs/} is searched for objects within the Stripe 82 area.  To remove duplicate objects from overlapping observations, only `primary' objects are retained, based on the appropriate flag in the {\sc PhotoObjAll} table.  NUV and FUV magnitudes and errors are returned as well as the coordinates of the objects.  We follow the approach of \citet{Salim:2007rv}: we use the {\tt calibrated magnitudes} which are the flux through an elliptical aperture scaled to twice the Kron radius for each source ({\sc MAG\_AUTO} from {\sc SExtractor}, \citealt{sex}) and so are approximately total magnitudes. Object detection is run separately in the NUV and FUV bands.  If detections are present in both bands, we use the elliptical aperture determined from the NUV image to determine the FUV flux ({\tt fuv\_ncat\_flux}), due to the typically greater sensitivity in the NUV.  If a FUV detection is present, but not NUV, then we take the aperture parameters from the FUV image ({\tt nuv\_fcat\_flux}).  Cross-matching is performed by associating {\it GALEX} objects (within a 0.6 degree radius of the pointing centre) with the closest positional match in the SDSS data within 4\arcsec.  This method should produce a reasonable cross-matched catalogue (e.g., \citealt{Salim:2007rv}, \citealt{Budavari:2009tg}).  We add 0.052 and 0.026 mag to the FUV and NUV magnitude errors respectively to account for calibration uncertainty \citep{Salim:2007rv}.  Galactic extinction is corrected following the prescription of \citet{Wyder:2007kn}.  The {\it GALEX} imaging covers 90\% of the 275 deg$^2$ SDSS Stripe 82 area.

\section{Measures of star formation rate}
\label{sec:emsfr}

In this section, we present the methods to derive the default measurements of SFR using: detailed fits of spectrophotometric data to stellar evolutionary synthesis and photoionsation models; and simple scaling relations between the luminosity of the \ha~and \oii~emission lines, and the luminosity of the ultraviolet continuum.  As described in the introduction, these SFR proxies have different limitations and sensitivities and our goal is to compare the accuracy with which these measure the total SFRD of the same volume of the local Universe.

\subsection{\b04 SFR$_{\rm Tot}$}
\label{sec:sfrtot}

\citet{Brinchmann:2004ct}, hereafter \b04, developed a method involving detailed modelling for deriving SFRs from SDSS spectra.  Two different approaches are used depending on whether the galaxy is classified as star-forming or exhibits an AGN component, based on the BPT diagram.  

For star-forming galaxies, the approach is to fit the spectrum using a suite of \citet{Bruzual:2003de} models to fit the continuum emission and also to naturally account for absorption by weak metal lines and Balmer lines.  Emission lines are then fitted using the models of  \citet{Charlot:2001gj} which model the emission lines of integrated spectra including a physically motivated model for dust attenuation.  By fitting to grids of such models and calculating likelihoods using a Bayesian approach, \b04~hence derive best fit values and associated probability density functions (PDFs) for various parameters for each galaxy such as SFR, metallicity, total dust attenuation, etc. See \b04 and references therein for details.  This fitting technique largely depends on the luminosity of the \ha~line, but, in practice, is estimated using all available emission lines. 

For galaxies which are either of too low S/N to use the emission line fitting technique, or for which the emission is from sources other than star-formation (i.e., AGN), an alternate approach is used.  The relation between the D4000 spectral index and the specific star-formation rate (SFR/$M_*$) is calculated for the {\it star-forming} sample just described, and this relation is then used to infer a star-formation rate for the remaining objects, based on a measurement of their D4000.  

In addition, \b04 adopt an empirical, colour-dependent aperture correction which attempts to correct for light missed by the fibre using the resolved colour from SDSS photometry.  Throughout this work, we adopt a simple aperture correction as described in \S\ref{sec:linefluxes} for all our emission line measurements.  We refer to the \b04~SFR derived from either the emission line fitting ($SFR_e$) or D4000 ($SFR_d$) converted to a total SFR in this way as \sfrtot.

The advantage of the \b04 method is that it allows detailed error estimates to be made by propagating full likelihood functions for various parameters over which the models are marginalised.  The disadvantage is that the estimated parameters are, by their very definition, highly model-dependent and also that this technique is not currently widely used, particularly at high redshift where spectra are typically of much lower S/N.  Nonetheless, we are interested in exploring how these complex estimates of SFR compare with more traditional measures.  Several recent works have used the \b04~SFR estimates as a reference against which to compare other estimates (e.g., \citealt{Treyer:2007gm}, \citealt{Argence:2009yg}), and here we do likewise.

\subsection{\ha$\lambda$6563~SFR}

Taking the \ha~line flux from the MPA catalogue (which has been corrected for absorption by the continuum fitting procedure used), the SFR can be estimated from the \ha~luminosity using 
\begin{equation}
\frac{SFR}{\mathrm{M_{\odot}\,yr^{-1}}} = \frac{10^{0.4 A_{\mathrm{H} \alpha}}}{1.5}
\frac{L(\mathrm{H}\alpha)}{1.27 \times 10^{41} \mathrm{\,erg\,s^{-1}}}
\label{eqn:sfr_ha_corr}
\end{equation}
The conversion factor corresponds to the Salpeter IMF calibration of
\citet{Kennicutt:1998pa} multiplied by 1.5 to convert to the Kroupa (2001)
IMF (e.g., B04).  This factor is similar to that obtained from the PEGASE stellar population code \citep{Fioc:1997cl} for solar metallicity, although in reality this factor
depends on metallicity and the extinction of Lyman continuum photons
before they ionise any nebular gas. The additional coefficient assumes $A_{H\alpha}$ mag of extinction at \ha.

We estimate the value of extinction at \ha~on a galaxy-by-galaxy basis using the observed Balmer decrement, the ratio of the fluxes, $f_{H\alpha}/f_{H\beta}$.  We assume an intrinsic Balmer decrement of 2.85 \citep{Osterbrock:1989yf} and  the extinction law of \citet{Seaton:1979or}, renormalised to $R_V=A_V/E(B-V)=3.1$, as is commonly done (e.g., \citealt{Jansen:2001qx}).  This gives 

\begin{equation}
\label{eqn:bd}
A_{H\alpha} = \frac{2.5}{k_{H\beta}/k_{H\alpha}-1} \log \left( \frac{1}{2.85} \frac{f_{H\alpha}}{f_{H\beta}} \right),
\end{equation}
where the coefficients $k_{H\beta}/k_{H\alpha}$ depend on the extinction law assumed, $k(\lambda)$, and $k_{H\beta}/k_{H\alpha} =1.48$ for the Seaton law (or similarly, 1.45 for the \citealt{Cardelli:1989oj} law).  The Balmer decrement-corrected \ha~SFR (which we shall refer to as \ha$_{corr}$) is commonly used as the preferred estimate of the SFR. The efficiency factor in converting ionizing photons/SFR into \ha~line luminosity varies weakly with metallicity \citep{Charlot:2001gj} (which may be linked to a galaxy's stellar mass via the mass--metallicity relation), such that metal-rich stars produce lower \ha~fluxes for a given SFR, and thus the \ha~flux would overestimate the SFR for high metallicity galaxies.  This effect is only a factor of $\sim$1.6 over three decades in mass, though (\b04, fig. 7).  

We adopt \ha$_{corr}$ as our preferred estimate of SFR, but will later show how this compares with the more detailed modelling to derive the SFR.

\subsection{u-band SFR}
The ultraviolet continuum can be used to estimate the luminosity from young stars and hence the SFR.  However, care must be taken to account for the contribution from older stellar populations to the ultraviolet light.  As a simple method for accounting for the old stellar populations, we reject galaxies on the red sequence.  This should only reject a minimal amount of the total SFRD in dust-reddened star-forming galaxies ($\sim$5\%, \citealt{Salim:2007rv}).  We derive a calibration, using PEGASE, between the UV luminosity and the star-formation rate (assuming a constant SFR over 1Gyr), for a Kroupa IMF around solar metallicity (although the calibration is similar for sub-solar metallicities), where UV can refer to any broad band filter in the UV regime (e.g., $u$-band, {\it GALEX} $NUV$, $FUV$, etc.).  The shape of the UV continuum ($\sim1500-3500\AA$) for a constant-SFR galaxy is approximately flat in $f_{\nu}$ and thus the coefficient linking UV luminosity and SFR is relatively independent of the filter definition.  For simplicity, we adopt an average relation of

\begin{equation}
\frac{SFR_{\mathrm{UV}}}{\mathrm{M_{\odot}\,yr^{-1}}} = 10^{0.4 A_{\mathrm{UV}}}
\frac{L_{\mathrm{UV}}}{1.4 \times 10^{21} \mathrm{\,W\,Hz^{-1}}}
\label{eqn:sfr_u}
\end{equation}
where $A_{UV}$ refers to the extinction in the UV passband chosen. We discuss different methods of estimating $A_{UV}$ in \S\ref{sec:comp_inds}.

Observed magnitudes are k-corrected to $z=0.1$, again using {\sc kcorrect} on the $ugriz$ photometry.  To remove the contribution from likely non-star forming galaxies, galaxies are separated into red and blue populations using the criterion $u-g = -0.033M_u+0.75$ in rest-frame magnitudes\footnote{The division in \citet{Prescott:2009vz} is given in rest-frame magnitudes, i.e. z$=$0, so although we transform luminosites to z$=$0.1 to minimise the uncertainty in the $k$-correction for SFRs, we $k$-correct to z$=$0 when separating red and blue galaxies} \citep{Prescott:2009vz}, and only the blue (star forming) population is considered when measuring SFRs.  
 
To examine the fraction of $u$-band flux from various stellar ages within the blue galaxy population, we consider a continuous-SFR 10 Gyr model from PEGASE \citep{Fioc:1997cl}, Fig.~\ref{fig:timescales}.  Taking a single model that has had a 10 Gyr constant-SFR at solar metallicity, the curves show the fractional contribution to the total flux in a given passband for bins in stellar age. For example, even though the SFR has been constant for 10 Gyr, nearly all the Lyman continuum ($<$91.2nm) radiation comes from $<$10 Myr stellar ages. Thus, the \ha~and \oii~indicators are sensitive to star-formation on these timescales. The FUV (shown at $\sim$150nm) is sensitive to star-formation on somewhat longer timescales (mostly $\sim$10 Myrs, but up to $\sim$1 Gyr). Now, the $u$-band ($\sim$325nm) exhibits a longer tail out to much older stellar ages, since more evolved stellar populations (if present) can contaminate the $u$-band flux. By applying a colour cut in order to reject red galaxies, we have removed galaxies which will have an even higher contribution from stars older than $\gsim$1 Gyr than for the continuous-SFR population. This leaves us with blue galaxies for which the dominant contribution to the $u$-band light is from recent star-formation. For a quasi-continuous SFR (the blue population), about 10\% of the $u$-band flux comes from stars older than 1Gyr for which we do not apply a correction.  The crucial point when we come to compare the cosmic average of the various indicators later is that, although the emission lines are sensitive to star-formation on much shorter timescales ($\leq$10Myr) than the UV estimators $\lsim1$ Gyr, both these timescales are short compared with the timescale over which the global star formation rate of the Universe is changing (e-folding time $\sim$3-4 Gyr, e.g., \citealt{Baldry:2003ue}). Thus, even though individual galaxies may have longer or shorter timescales for star-formation than $\sim$1 Gyr, say (and thus show differences in their individual SFRs estimated from \ha~versus UV, for example), {\it the cosmic average} will show the same behaviour in all the different indicators, since the timescales over which these are sensitive are short compared with the timescale on which the global average is changing. 

\begin{figure}
{\centering
\includegraphics[width=85mm]{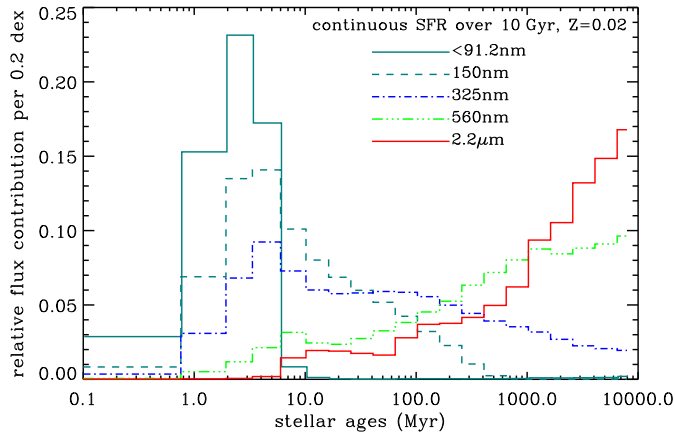}
\caption{The flux contribution versus stellar age for a star-forming galaxy that
has had a constant SFR over 10 Gyr. Model spectra were obtained from PEGASE with no extinction, and the different line styles (as indicated in the key) indicate the results for different passbands. The UV passbands 150nm (dashed line) and 325nm (dot-dashed line) are sensitive to star-formation on short timescales (ages $\lsim1$ Gyr) but the latter exhibits a tail toward longer timescales, meaning that more evolved stellar populations can contaminate SFR estimates using 325nm flux alone. See text for details. Note also that the Lyman continuum flux (solid blue-green line) responsible for producing the H$\alpha$ emission line is sensitive only to ages $\lsim$10 Myr.
\label{fig:timescales}
}
}
\end{figure}

\subsection{FUV SFR}

Observed FUV$-$band photometry from {\it GALEX} should be a better proxy for the rest-frame ultraviolet continuum produced by young stars than the observed $u$-band, since it is less affected by more evolved stellar populations.  The disadvantage of the FUV is that it is more susceptible to the effects of dust extinction than the redder, $u$-band luminosity.  To estimate FUV luminosities, $k$-corrections to $z=0.1$ are derived from FUV, NUV, $ugriz$ photometry using the {\sc galex\_kcorrect} routine of {\sc kcorrect}.  FUV luminosity is then converted to SFR using Eqn.~\ref{eqn:sfr_u}.   

Again, we only consider blue galaxies when measuring SFRs from the FUV, this time using the rest-frame $(NUV-r)<4$ colour criterion of \citet{Salim:2007rv}.  According to those authors, 95\% of star-formation takes place blueward of this colour cut, and thus this is an observationally efficient way to select the galaxies of interest whilst eliminating possible AGN contaminants.  \citet{Salim:2007rv} provide a relation to estimate the FUV extinction using FUV-NUV colours (their eq.~6) and we adopt this to produce an extinction-corrected estimated of the FUV SFR. 

\subsubsection{Dust corrections to the UV SFRs}
\label{sec:dustcorr}

The dust correction to the UV ($u$-band and FUV) SFRs is one of the largest uncertainties. As described above, for each passband we will use three different methods: a) assuming a fixed constant extinction of A(\ha)$=$1 mag, b) a value scaled from the Balmer decrement and c) for the FUV, a correction based on the UV slope (NUV-FUV colour); and for the $u$-band, a correction which depends on the SFR. The first two methods require the knowledge of the relative extinction between that of the \ha~emission line and the UV wavelength of interest for the stellar continuum. One commonly used prescription is that of \citet{Calzetti:2000sp}. This is derived from starburst galaxies and may not directly apply to less vigorously star-forming galaxies \citep{Buat:2002fy}. The standard implementation of the \citet{Calzetti:2000sp} method requires the use of a Milky Way-like screen law (e.g., \citealt{Cardelli:1989oj}) to estimate the reddening suffered by the emission line. The obscuration curve of \citet{Calzetti:2000sp} is then used to estimate the reddening experienced by the stellar continuum at the wavelength of interest (here, 3220\AA~and 1390\AA~for the z$=$0.1 $u$-band and FUV respectively), with an additional factor of 0.44 to account for the greater reddening experienced by the emission lines. This leads to relations of $A(u) = 1.19 A(H{\alpha})$ and $A(FUV) = 1.96 A(H{\alpha})$. Now, the reason for using the two different obscuration curves (foreground screen for lines and an embedded geometry for the stars) for a starburst galaxy is sketched out in \citet{Calzetti:2001dk}; but is likely an overestimate for more modestly star-forming galaxies. In order to attempt to mitigate this effect, we use the \citet{Calzetti:2000sp} obscuration curve for both the stars and the emission lines (but again use the factor of 0.44 to convert between the relative reddening in the two). This yields relations of 
\begin{equation}
A(u) = 0.87 A(H{\alpha})
\label{eqn:extu}
\end{equation}
 and 
 \begin{equation}
 A(FUV) = 1.43 A(H{\alpha}).
\label{eqn:extfuv}
 \end{equation}
 This approach is somewhat similar to the model of 
\citet{Charlot:2000zm} who use a single attenuation law for birth clouds and the diffuse ISM, with a different normalisation between the two. These numbers are also broadly in agreement with the observational result of \citet{Buat:2002fy} who found that the standard \citet{Calzetti:2000sp} prescription overestimates the UV extinction by $\sim$0.6 mag when applied to non-star bursting galaxies. 
We obtain broadly similar results ($A(u) = 0.94 A(H_{\alpha})$ and $A(FUV) = 1.55 A(H_{\alpha})$) if, instead of using the \citet{Calzetti:2000sp} obscuration curve to correct both the stars and the gas, we use only the Milky Way extinction curve of \citet{Cardelli:1989oj} for both (again using the factor of 0.44 to convert between stars and gas).

Regardless of the exact motivation, we will show that these scalings between extinction at \ha~and the UV (eqn.~\ref{eqn:extu} \& eqn.\ref{eqn:extfuv}) yield reasonably good agreement between the SFRD estimated in the UV and \ha~and so we adopt these as our default normalisations. Our SFRDs may straightforwardly be renormalised to assume a different dust prescription by simply multiplying the SFRD by $10^{0.4(old-new)}$ where $old$ represents the coefficient in eqn.~\ref{eqn:extu} or \ref{eqn:extfuv} and $new$ is the desired alternate coefficent. We will discuss the choice of extinction correction further in \S\ref{sec:comp_inds}.

\subsection{{\rm \oii}$\lambda$3727~SFR}
\label{sec:oii}

The SFR measured from \oii~luminosity may be estimated by scaling between the \oii~and \ha~luminosity  using

\begin{equation}
\label{eqn:sfr_oii}
\frac{SFR}{\mathrm{M_{\odot}\,yr^{-1}}} = 
\frac{10^{0.4 A_{\mathrm{H} \alpha}}}{1.5\,r_{\mathrm{lines}}}
\frac{L(\mathrm{[O\,II]})}{1.27 \times 10^{41} \mathrm{\,erg\,s^{-1}}}
\end{equation}
where  $r_{\rm lines}$ is the ratio of extinguished \oii~to \ha~flux.  In the absence of better information, a ratio of $\approx0.5$ is typically assumed (e.g., \citealt{Kennicutt:1998pa}).  We take this value of 0.5 and assume 1 mag of extinction at \ha~(as is commonly done) as our nominal \oii~SFR.  This gives a nominal SFR relation, $SFR_0/(M_\odot yr^{-1}) = L({\rm [O{\sc II}]})/(3.80 \times 10^{40} erg~s^{-1})$.

It is well known that the ratio, $r_{\rm lines}$, depends on metallicity and ionisation parameter (e.g., \citealt{Kewley:2004vs} and references therein).  As a default model, we assume a fixed conversion between \oii~luminosity and SFR, as is often done.  In \S\ref{sec:empoii} we will consider empirical corrections which attempt to account for this (i.e., without needing to assume a fixed $r_{\rm lines}$).

\section{Empirical correction to the \oii~SFR}
\label{sec:empoii}

\begin{figure*}
{\centering
\includegraphics[width=140mm]{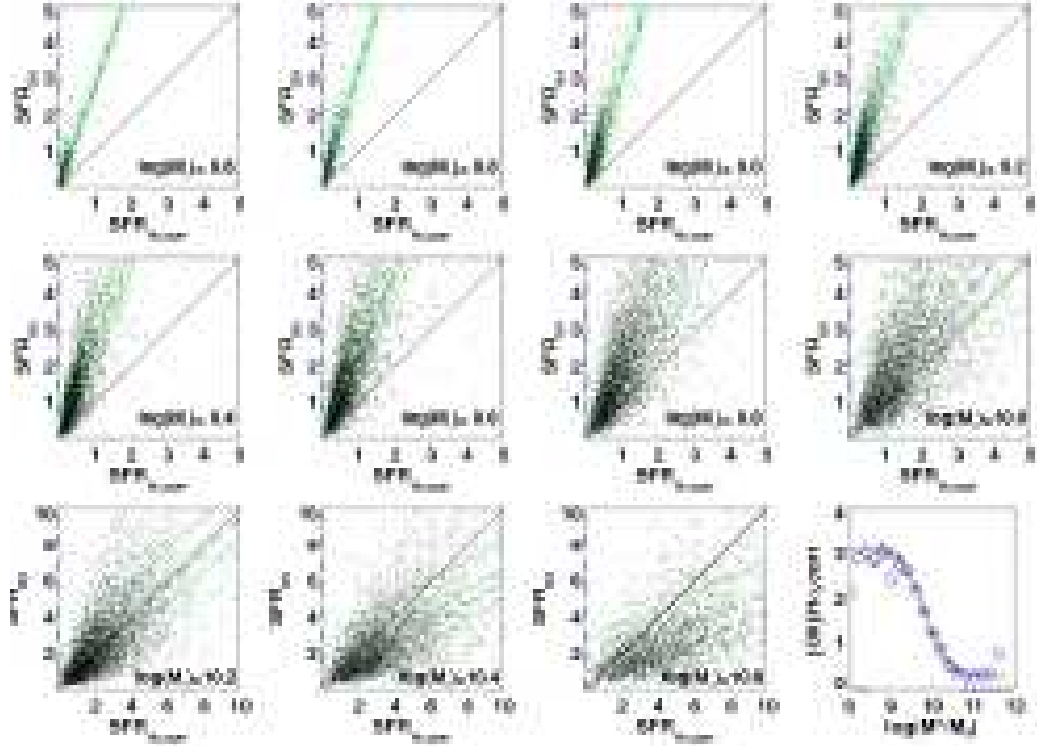}
\caption{Comparison of Balmer decrement-corrected \ha~SFR against constant extinction \oii~SFR rate separated into bins of stellar mass.  Note: fit is performed over full range of SFRs, but plots are scaled to show most populated regions for clarity. Best fit lines with 1$\sigma$ uncertainties are shown in green.  Dashed blue line indicates the slope adopted in the best-fit fitting function in each mass bin.  The solid black line indicates equality, for reference. The final panel shows the best fit ratio of $SFR_{[OII]}$/$SFR_{H\alpha}$ as a function of stellar mass (black points with error bars); the green line is the adopted tanh correction.  For reference, blue diamonds indicate the ratio of the total SFRD from \oii~to \sfrtot~(Fig.~\ref{fig:sfrd_lines}), but this is not used in the fit.  See text for details.
\label{fig:oii_ha_mass}
}
}
\end{figure*}

The \oii~luminosity from the \oii$\lambda\lambda3726, 3729$ doublet is less-directly coupled to the ionising flux from young stars than \ha, and thus suffers larger uncertainties as a SFR indicator.  As mentioned above, many of the systematic uncertainties in the SFR indicators may be related to stellar mass (e.g., through the mass--metallicity relation).  So, it is of interest to examine how the relation between \oii~SFR (estimated simply from \oii~luminosity with a constant conversion to SFR) compares with a more direct SFR indicator, on a galaxy-by-galaxy basis, as a function of stellar mass.  Fig.~\ref{fig:oii_ha_mass} shows \oii~SFR versus the Balmer decrement-corrected \ha~SFR.  For this comparison, only galaxies identified as belonging to the star-forming class by the BPT classification described in \S\ref{sec:linefluxes} are considered.  In each mass bin, a robust linear regression fit (required to pass through the origin) is performed rejecting outliers.  The best fit and its 1$\sigma$ uncertainty are indicated.  It is clear from the figure that the slope of the relation is a strong function of mass.  In order to model this dependence, the best-fit slope and its uncertainty in each mass bin is fitted with a tanh function of mass (shown in the final panel of Fig.~\ref{fig:oii_ha_mass}).  The tanh function effectively captures the shape of this relation, being approximately horizontal at high and low masses and switching between these two levels in an approximately linear fashion at intermediate masses. The empirically-corrected SFR, $SFR_{emp,corr}$ is obtained from the nominal \oii-SFR (Eqn~\ref{eqn:sfr_oii}), $SFR_0$,

\begin{equation}
SFR_{emp,corr} = \frac{SFR_0}{a\,\tanh[(x-b)/c] +d},
\label{eqn:oiipoly}
\end{equation} 
where $x=$ \ms~and $a=-1.424$, $b=9.827$, $c=0.572$, and $d=1.700$, and $SFR_0/(M_\odot yr^{-1}) = L({\rm [O{\sc II}]})/(2.53 \times 10^{40} erg~s^{-1})$ from above. This. should give the most reliable estimate of SFR from \oii~luminosity.  The correction between \oii~and \ha$_{corr}$ SFR encodes all the information about the metallicity and dust dependence of \oii~luminosity on SFR in each mass bin.  In principle, this can now be used to correct \oii~luminosity to SFR in higher redshift. If this correction is used at higher redshift, this implictly assumes that the mass-dependence of extinction, metallicity, etc. (which are all encoded in the correction) behave in the same way at higher redshift as locally.  We note that this (possibly incorrect) assumption is commonly used anyway at higher redshift when locally-derived relations (e.g., Eqn.~\ref{eqn:sfr_oii}) are applied to higher redshift galaxies.  We will consider this point further in \S~\ref{sec:highz}, but for the moment note that the factor most likely to evolve between here and the higher redshift Universe is likely to be the amount of dust in a galaxy of given mass.  Naively, one might assume that the dust content of a galaxy would increase from high to low redshift, as successive generations of stars pollute their host galaxy.  However, it is equally possible to make simple arguments for the contrary evolution based on dust geometry, for example.  To simplify comparison with future work, Fig.~\ref{fig:avedust} shows the amount of dust extinction, as measured from the Balmer decrement, as a function of stellar mass.  The solid line shows a fit to the mass-dependent extinction, which may be modelled as 

\begin{equation}
A_{\mathrm{H} \alpha} = a + b \log(M_{*} / \mathrm{M_{\odot}}) + 
  c [\log(M_{*} / \mathrm{M_{\odot}})]^2
\label{eqn:massdep}
\end{equation}
with the values $a=51.201$, $b=-11.199$, $c=0.615$ and setting the fit to a constant value below \ms$\le$9.0.  For reference, the dashed line shows the average mass dependence of extinction found by \b04, taking the value from the peak of their likelihood functions in each mass bin from their fig.~6.  The coefficients are approximately $a=-3.39, b=0.46~(c=0)$.  Their relation is different from ours since we use the Balmer decrement whereas they estimate the dust by simultaneous fits to many parameters in the \citet{Charlot:2001gj} models. 
These models include a component of diffuse nebular emission which has a lower effective temperature than the canonical value assumed for HII regions and, in effect, leads to the adoption of a non-constant ratio for the Balmer decrement (J. Brinchmann, priv. comm.). This is phrased as ``metallicity dependence of the Case B \ha/H$\beta$ ratio" in \b04. It can be seen from Fig.~\ref{fig:avedust} that, for low mass (\ms$\sim$9) galaxies, allowing for a non-constant Balmer decrement ratio leads to extinction estimates approximately twice as large as those inferred from our constant Balmer ratio estimate. When estimating SFR, this is almost precisely cancelled by a corresponding variation in conversion factor between \ha~luminosity and SFR in the \citet{Charlot:2001gj} models (fig.~7, \b04). Given the uncertainties in fitting such a diffuse component, we prefer to adopt the more traditional constant Balmer decrement and constant conversion from \ha~luminosity to SFR. \b04 showed that the these two approaches give very similar estimates of SFR (and we will also show that this is true when we compare SFRDs derived from the different indicators).  Furthermore, we also show in Fig.~\ref{fig:avedust} (larger filled circles) an alternate estimate of extinction based on FIR measures (discussed in  \S\ref{sec:othemp}) which can be seen to be in better agreement with that derived from our constant Balmer decrement estimate than the variable ratio model fits.

These various parameterisations of how dust extinction varies with mass will allow the reader to remove the dust term from our empirical correction and substitute an alternate correction, if desired.  In addition, Fig.~\ref{fig:avedust} in conjunction with Fig.~\ref{fig:oii_ha_mass} shows how much of the systematic offset of \oii~to~\ha~SFR is due to the effect of dust.

\begin{figure}
{\centering
\includegraphics[width=80mm]{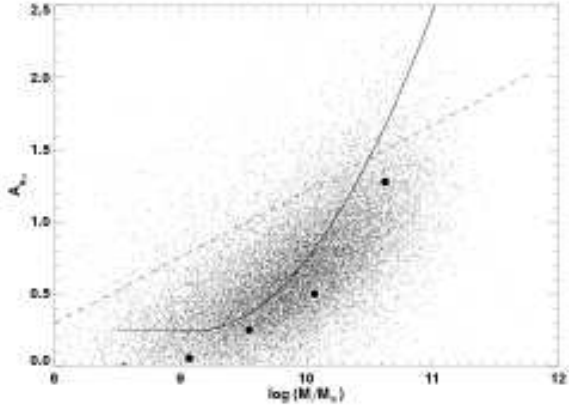}
\caption{Extinction in the \ha~line (estimated from the Balmer decrement) as a function of stellar mass.  Points show measurements for each galaxy, and the solid line is our quadratic fit described in the text.  Dashed line shows the relation inferred from \b04~(which uses fits to the models of \citealt{Charlot:2001gj} to attempt to estimate the dust). Larger filled circles show extinction scaled from IR estimates, discussed in \S\ref{sec:othemp}.
\label{fig:avedust}
}
}
\end{figure}

\section{Emission line luminosity functions}

Before proceeding to more detailed studies using SFRs based on our \oii~and \ha~measurements, it may be instructive to compare the distributions of our measurements with those of other surveys. The simplest useful statistic employed by many works is that of the luminosity function. In order to more easily compare with other works we use our observed quantities directly. These may simply be rescaled using the relations given above, if the reader desires to see these in terms of SFR functions, rather than observed emission line luminosity. Another useful result of such a comparison is that it will allow us to confirm the depth of our emission line flux limit, by examining the faintest lines visible. 

We follow the method of \citet{Zhu:2008iv} and use a $1/V_{max}$ technique to estimate the number of objects in bins of log luminosity (see also \S\ref{sec:vmax}).  Error bars are estimated using scaled Poissonian distributions, as described in \citet{Zhu:2008iv}. The resulting LFs are listed in Table \ref{tab:lf} and plotted in Fig.~\ref{fig:lf}. Filled red circles show the \ha~LF and open blue diamonds show that for \oii.  Neither LF is well fitted by a Schechter function, which is the normally-employed fitting function for galaxy LFs.  As noted by \citet{Zhu:2008iv} at z$\sim$1, a better function to use is a double power law.  They found that the higher redshift \oii~LFs are well-described by a power law with a break around $\log L_{OII}\sim42$ with a slope, $\alpha\sim-3$ ($dN/dL \propto L^\alpha$) at higher luminosities, and $\alpha\sim-1.3$ faintward of the break.  Fitting a similar function, we find that the \ha~LF is well-described by a function with $\alpha_{bright}=-2.48$, $\alpha_{faint}=-1.27$ and a turnover luminosity, $\log L_{TO}=41.33$.  For the \oii~LF, these values are $\alpha_{bright}=-2.83$, $\alpha_{faint}=-1.67$ and $\log L_{TO}=41.03$.  The dotted vertical lines \ha~and \oii~(red and blue respectively) show the median luminosities at which the S/N falls to 2.5 (our selection limit).  

\begin{figure}
{\centering
\includegraphics[width=80mm]{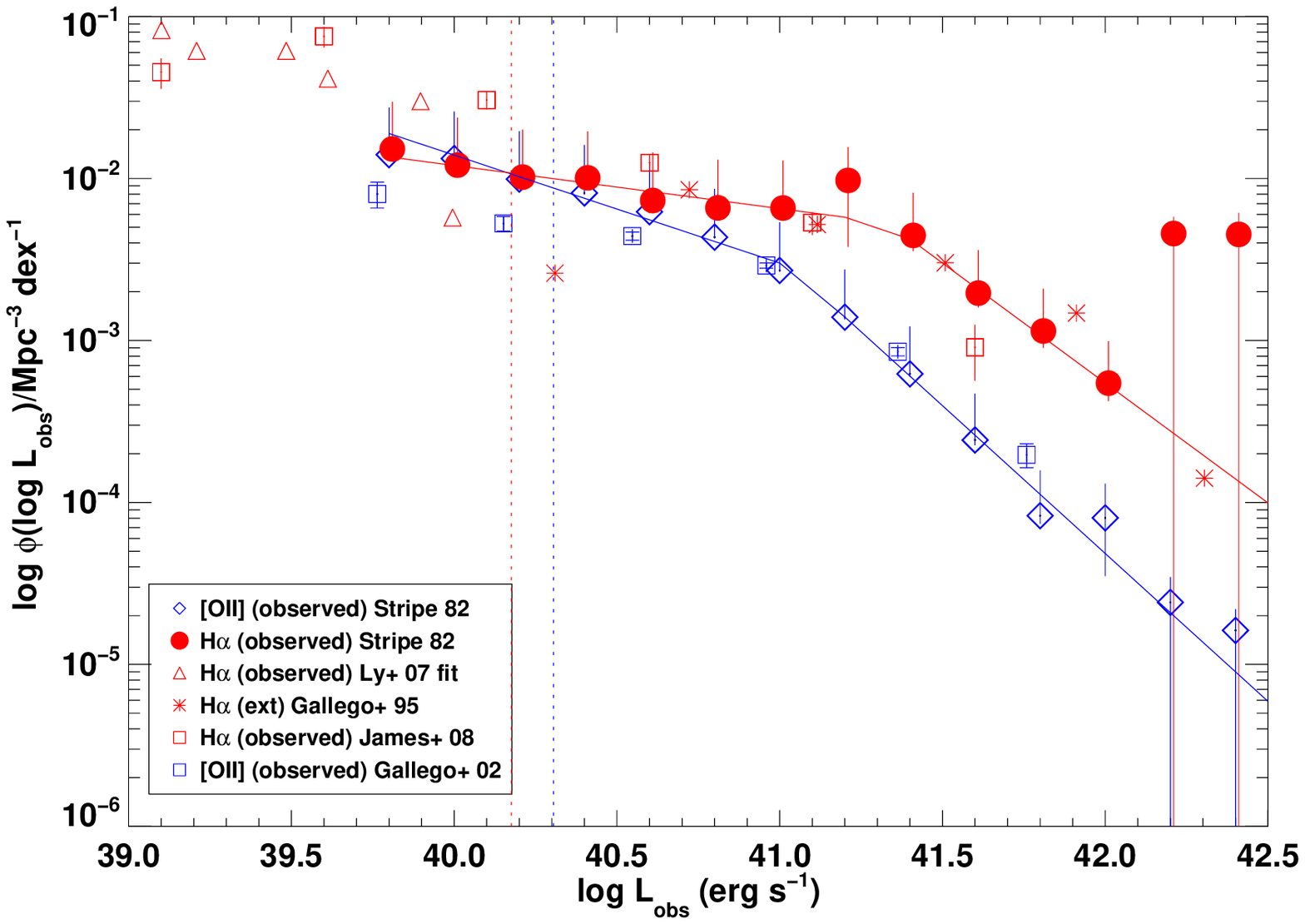}
\caption{Luminosity functions for \ha~and \oii~in Stripe 82.  Red and blue lines with error bars show observed (i.e., no correction for extinction) luminosity functions for \ha~and \oii~respectively.  For comparison, observed \oii~(z$\lsim$0.045) from \citet{Gallego:2002eu} (blue squares);  extinction corrected \ha~(z$\lsim$0.045) from \citet{Gallego:1995la} (red asterisks); non-extinction corrected \ha~from the $z < 0.01$ narrowband survey of \citep{James:2008wd} (P. James, priv. comm.); and observed \ha~from the z$\sim$0.08 narrowband survey of Ly et al. (2007) (red triangles) are shown.  Double power law fits to the Stripe 82 data are overplotted.  Dotted vertical lines (red: \ha; blue: \oii) illustrate luminosities corresponding to SFRs$\approx0.1 M_\odot yr^{-1}$ for the simple constant conversions.  See text for discussion. 
\label{fig:lf}
}
}
\end{figure}

\begin{table}
\caption{Luminosity functions for \oii~and \ha~in Stripe 82, plotted in Fig.~\ref{fig:lf}.   Number densities, $\phi$, are observed values with no corrections for incompleteness or internal extinction.  Incompleteness will begin to set in fainter than $\log L \approx$40.3.  See text for details.
\label{tab:lf}
}
\begin{tabular}{lcc}
\hline
$\log L$ & $\phi_{OII}$  & $\phi_{H\alpha}$ \\
             & $(10^{-3} Mpc^{-3} dex^{-1})$ & $(10^{-3} Mpc^{-3} dex^{-1})$ \\
\hline
39.80 & $ 14.031^{+  0.670}_{- 13.391}$ & $ 15.219^{+  0.807}_{- 14.452}$ \\
40.00 & $ 13.292^{+  0.782}_{- 12.552}$ & $ 12.106^{+  0.471}_{- 11.653}$ \\
40.20 & $  9.922^{+  0.248}_{-  9.679}$ & $ 10.201^{+  0.371}_{-  9.843}$ \\
40.40 & $  8.141^{+  0.186}_{-  7.960}$ & $ 10.097^{+  0.703}_{-  9.439}$ \\
40.60 & $  6.230^{+  0.119}_{-  6.113}$ & $  7.317^{+  0.148}_{-  7.172}$ \\
40.80 & $  4.343^{+  0.079}_{-  4.265}$ & $  6.587^{+  0.113}_{-  6.476}$ \\
41.00 & $  2.707^{+  0.054}_{-  2.653}$ & $  6.573^{+  0.248}_{-  6.334}$ \\
41.20 & $  1.393^{+  0.047}_{-  1.347}$ & $  9.734^{+  5.944}_{-  5.868}$ \\
41.40 & $  0.622^{+  0.027}_{-  0.596}$ & $  4.455^{+  0.895}_{-  3.703}$ \\
41.60 & $  0.243^{+  0.017}_{-  0.227}$ & $  1.961^{+  0.361}_{-  1.654}$ \\
41.80 & $  0.083^{+  0.009}_{-  0.075}$ & $  1.144^{+  0.244}_{-  0.941}$ \\
42.00 & $  0.080^{+  0.045}_{-  0.050}$ & $  0.545^{+  0.123}_{-  0.444}$ \\
42.20 & $  0.024^{+  0.026}_{-  0.010}$ & $  4.573^{+  8.089}_{-  1.236}$ \\
42.40 & $  0.016^{+  0.022}_{-  0.006}$ & $  4.519^{+  6.083}_{-  1.631}$ \\
\hline
\end{tabular}
\end{table}

Local LFs from the literature are overplotted on the figure.  There are few recent \oii~LFs, since \ha~is the preferred SFR indicator locally, and is more easily accessible (at very low redshifts, few spectrographs have sufficient blue sensitivity to the observed wavelength of \oii).  The observed \oii~(i.e., uncorrected for extinction) LF from \citet{Gallego:2002eu} (converted to our cosmology) for a sample of $\sim$150 galaxies at z$\lsim$0.045 is shown as open blue squares.  This is in excellent agreement with the Stripe 82 measurement at the bright end.  The turnover occurs in the same place in both datasets.  However, the faint end slope appears somewhat shallower than ours in the  \citet{Gallego:2002eu} sample.  This may be a sign of incompleteness in the latter dataset, and the much greater size and uniformity of the SDSS data should be favoured.  \citet{Gallego:1995la} presented an \ha~LF from the same survey as their \oii~LF.  However, in this case they only published extinction-corrected \ha~luminosities, shown as red asterisks in Fig.~\ref{fig:lf}.  Even so, these are in excellent agreement with our observed (non-extinction corrected) LF (except for their faintest bin, which is likely again incomplete).  At lower luminosities, \citet{Ly:2007hc} published an \ha~LF based on narrowband imaging of the Subaru Deep Field  at z$\sim$0.08.   This was drawn from too small an area to constrain the bright end of the LF, however the faint end appears to have approximately the same slope ($\alpha \approx -1.6$) as ours, but with higher normalisation.  This may be due to cosmic variance, due to the small field size, or possibly some other systematic difference between the datasets based on narrowband imaging and flux-calibrated spectroscopy.  The narrowband \ha~imaging survey of \citet{James:2008wd} similarly shows a steeper faint end slope (fainter than our completeness limit) suggesting that this is indeed likely due to completeness effects.  Brighter than this, the \citet{James:2008wd} \ha~luminosity function is in good agreement with our measurements for all but the brightest point which disagrees at $\sim$2$\sigma$.  This could again be due to cosmic variance, due to the relatively small area, or to evolution, since this sample is selected to be $z<0.01$. 

All our results for \ha~and \oii~so far have considered galaxies with S/N$>$2.5 in the given emission line, since we require a minimum S/N in each emission line in order to assign an object an SFR from the emission line measurements.  An important question for understanding the SFRD measurement is how well an \oii-selected survey (or \ha-selected survey) mimics an SFR-selected survey.
The above comparison of the \oii~and \ha~LFs in Stripe 82 shows that the number densities of \ha-selected and \oii-selected objects at a given luminosity (SFR), close to the limiting luminosity, are comparable.   
 In order to more effectively study the effect of selection, we next consider directly the star formation rate density (SFRD) and its mass dependence.

\section{The star-formation rate density}

Whilst the comparison of SFR indicators on a galaxy-by-galaxy basis may be the most instructive method for understanding systematic differences between indicators, perhaps a more useful way for understanding results from large galaxy surveys is to examine the ensemble properties of galaxy populations such as the star formation rate density (SFRD).  
The advantage of studying this cosmically-averaged quantity is that the timescale sensitivities of the different indicators considered are all short compared with the timescale on which the cosmic average is changing, and this allows a fair comparison between the different indicators. Indeed, a comparison of UV SFR versus \ha~SFR for individual galaxies might be expected to show significant scatter, since these indicators are sensitive to star-formation on different timescales\footnote{We only use the galaxy-by-galaxy comparison to correct the \oii~SFR to that of \ha~since both these indicators are essentially sampling the same stars.}, but averaged over large galaxy populations, such differences are reduced. 

\subsection{Calculating SFRD}
\label{sec:vmax}

The star formation rate density in the $m^{th}$ mass bin, $\rho_{\rm SFRD}$, is calculated using the $1/V_{max}$ methodology (e.g., \citealt{Felten:1976em})

\begin{equation}
\rho_{SFR,m} = \frac{1}{\Delta \log M} \sum \frac{SFR_i}{V_{\mathrm{max},i}\,c_i}
\label{eqn:vmax}
\end{equation}
where $SFR_i$ is the SFR of the  $i^{th}$ galaxy, $V_{max,i}$  is the maximum volume out to which it could be observed in the survey, and the weighting, $c_i$, is the spectroscopic completeness described in \S\ref{sec:data}.  
If a galaxy is bright enough to be seen over the full redshift range ($0.032<z<0.20$) down to the magnitude limit of the survey ($r=19.5$), then it would be drawn from a volume of $1.5 \times 10^7$ Mpc$^3$. For galaxies which could not be seen out to the high redshift limit, it is necessary to calculate the redshift at which it would become too faint to be included.  This procedure involves knowing the spectral type of the galaxy in order to estimate its $k$-correction (at its maximum redshift, again using {\sc kcorrect}), and ultimately the maximum redshift (and hence $V_{max}$) must be calculated iteratively.

Fig.~\ref{fig:sfrd_lines} shows the SFRD as a function of stellar mass for the \b04~\sfrtot.  For reference, the \b04~estimate of the total SFRD from a larger area of SDSS (DR2, and in the redshift range $0.005<z<0.22$) is overplotted as the green dashed line.  Poisson uncertainties based on the number of galaxies in each mass bin are much smaller than the systematic uncertainties and so error bars for each individual method are omitted from the plot for clarity.

\begin{figure*}
{\centering
\includegraphics[width=85mm]{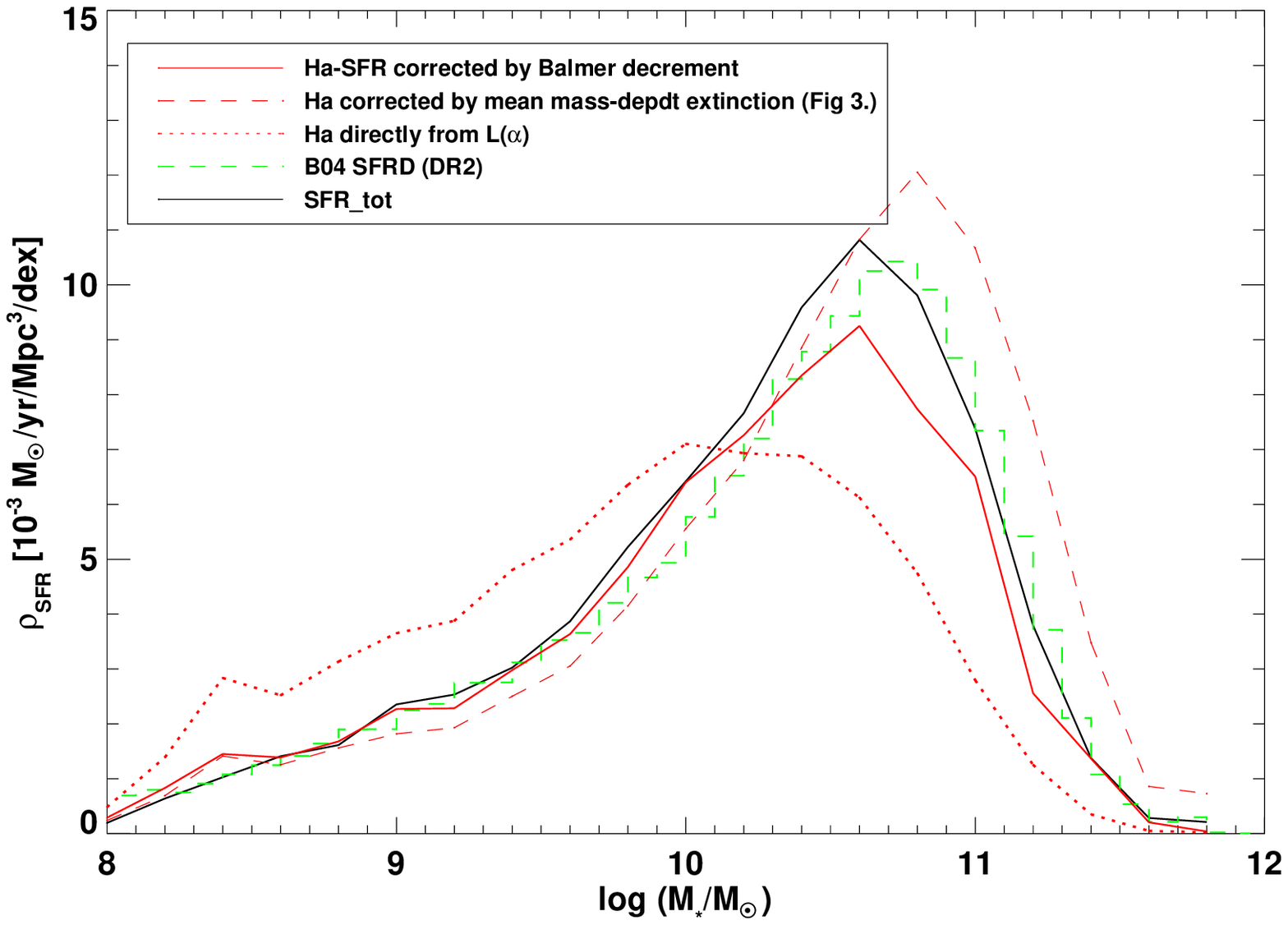}
\includegraphics[width=85mm]{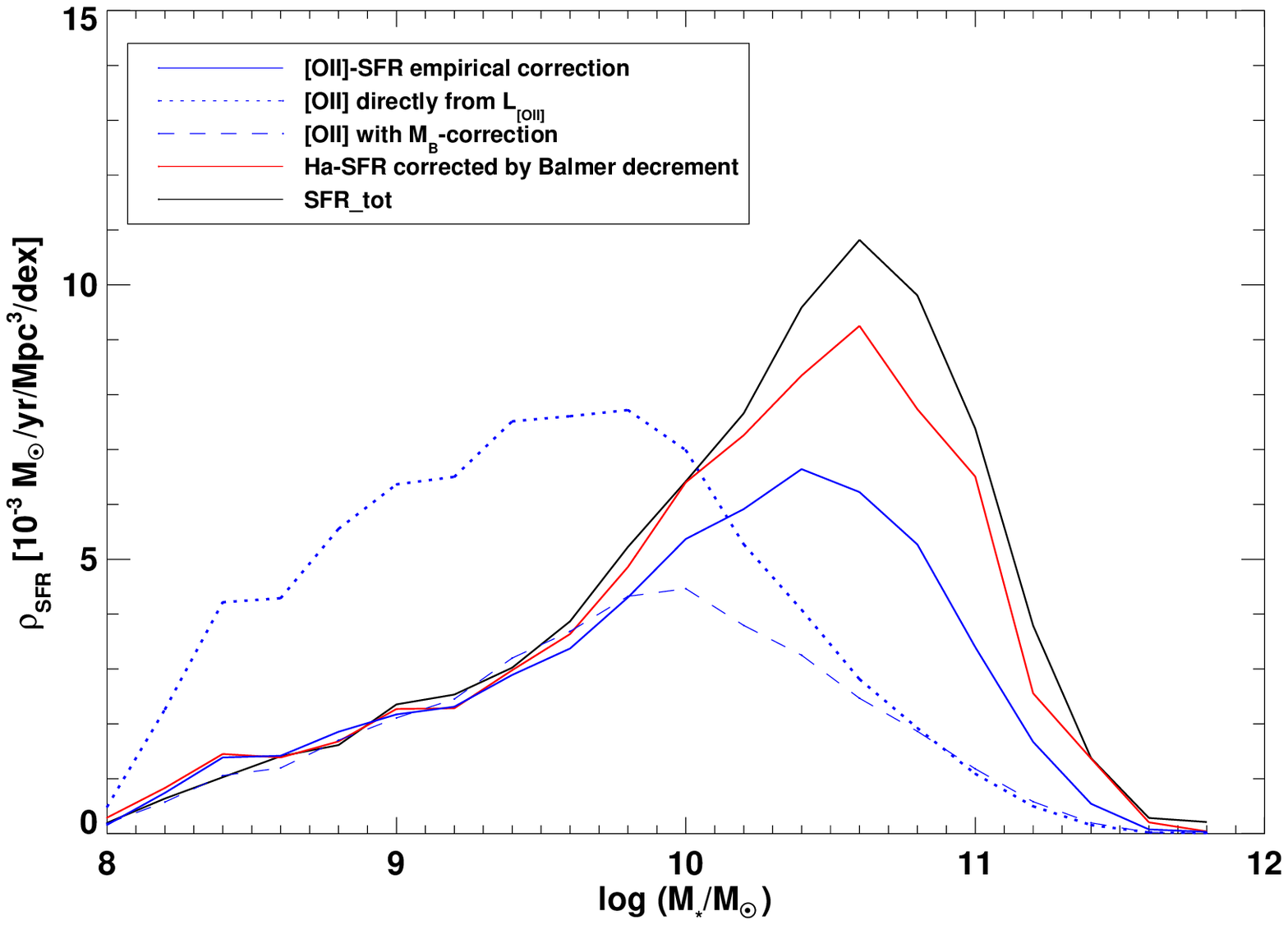}
\caption{The SFRD as a function of stellar mass for the different emission line measurements of SFR. The left panel shows rates based on \ha~(red lines): solid line is Balmer decrement-corrected (on a galaxy-by-galaxy basis) \ha; dotted line is for 1 mag of constant extinction at \ha~(in Eqn.~\ref{eqn:sfr_ha_corr}); dashed line assumes average mass-dependent extinction from Fig.~\ref{fig:avedust}.  The solid black line shows the SFRD calculated using \sfrtot~(the modelling technique of \b04); and the dashed green histogram shows the SFRD from \b04 for the larger SDSS DR2 sample.  The right panel shows the SFRD from \oii~assuming 1 mag of constant extinction at \ha~(dotted blue line) and that using the empirical correction to \oii~(solid line).  The empirically-corrected \oii~SFRD only underestimates the \ha~SFRD due to the \oii~flux limit (see also Fig.~\ref{fig:sfrd_thresh} and text for discussion).  For reference, \ha$_{corr}$ and \sfrtot~results are repeated from left panel.
\label{fig:sfrd_lines}
}
}
\end{figure*}

\begin{figure*}
{\centering
\includegraphics[width=140mm]{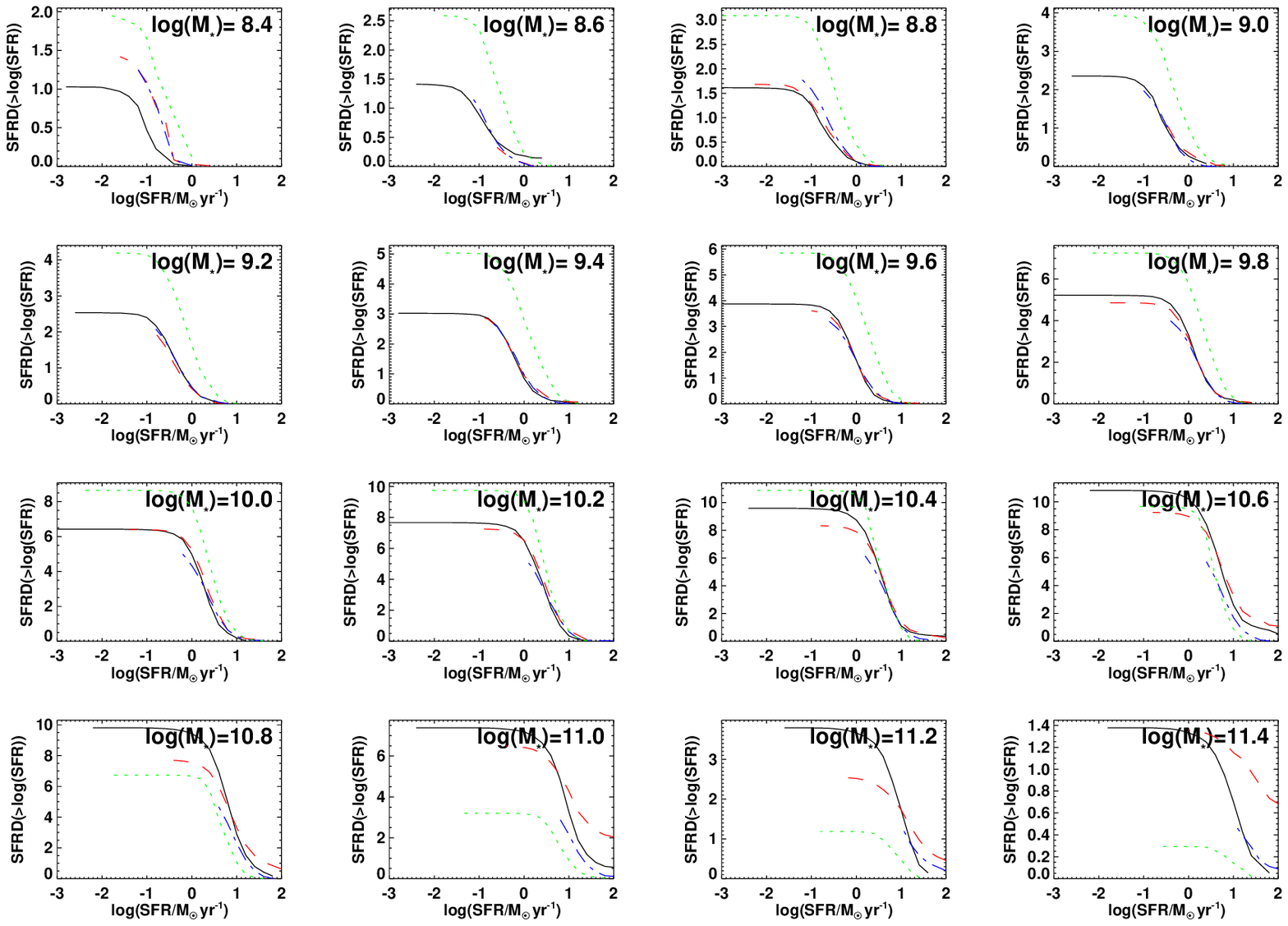}
\caption{The total SFRD as a function of limiting SFR in bins of stellar mass for \sfrtot~(black solid line), \ha$_{corr}$ (dashed red line), and \oii$_{emp,corr}$ (dot-dashed blue line).  Units of SFRD are $10^{-3} M_\odot\,yr^{-1} Mpc^{-3}$.  In stripe 82, the SFRDs based on \sfrtot~and \ha$_{corr}$ largely appear to have converged for a limiting SFR of $\log (SFR) \sim -1$.  However, the \oii~based SFRDs appear to not have converged in the highest mass bins, even below this limit.  See text for discussion.  
\label{fig:cumsfrd}
}
}
\end{figure*}

\begin{figure*}
{\centering
\includegraphics[width=85mm]{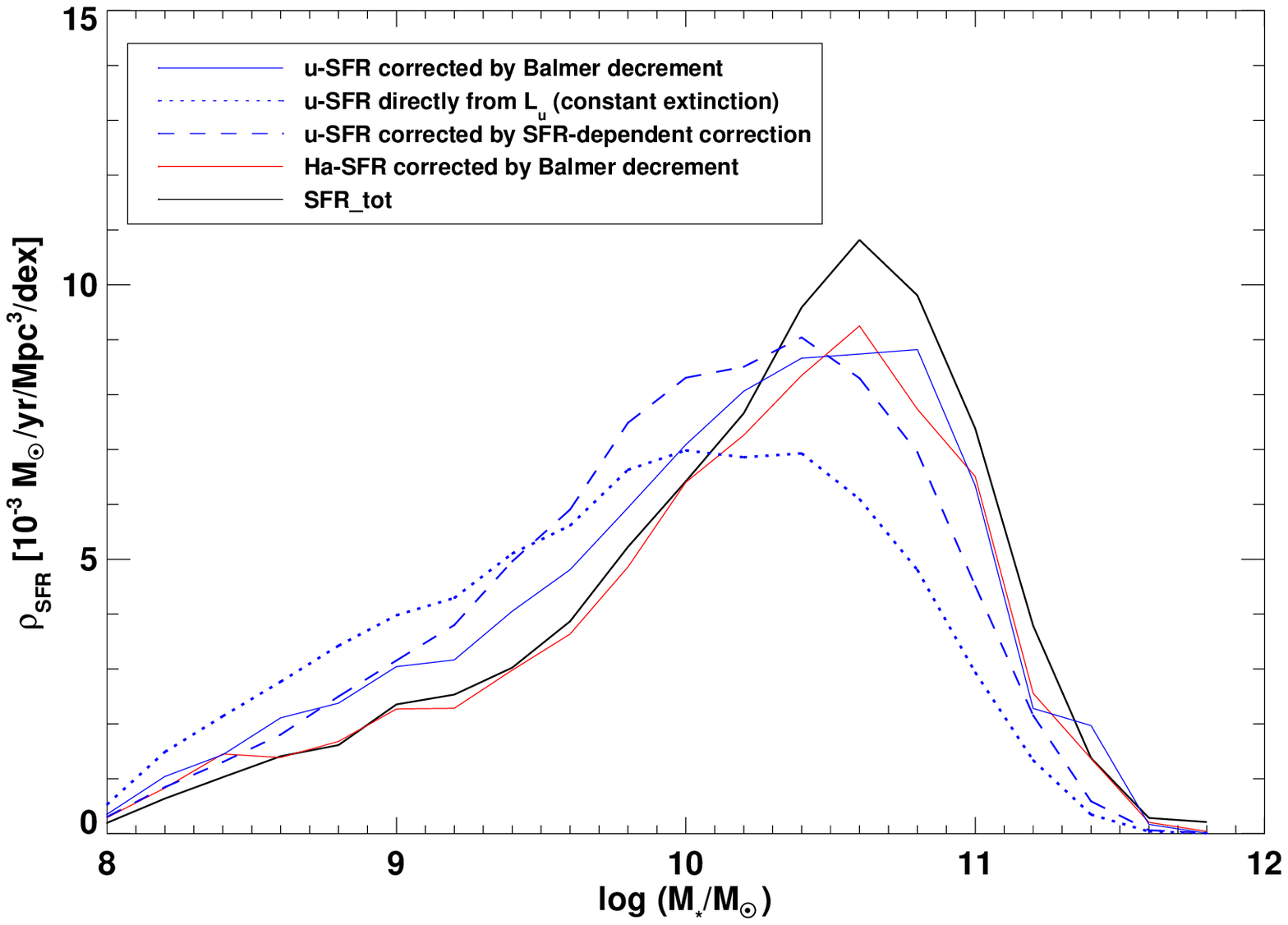}
\includegraphics[width=85mm]{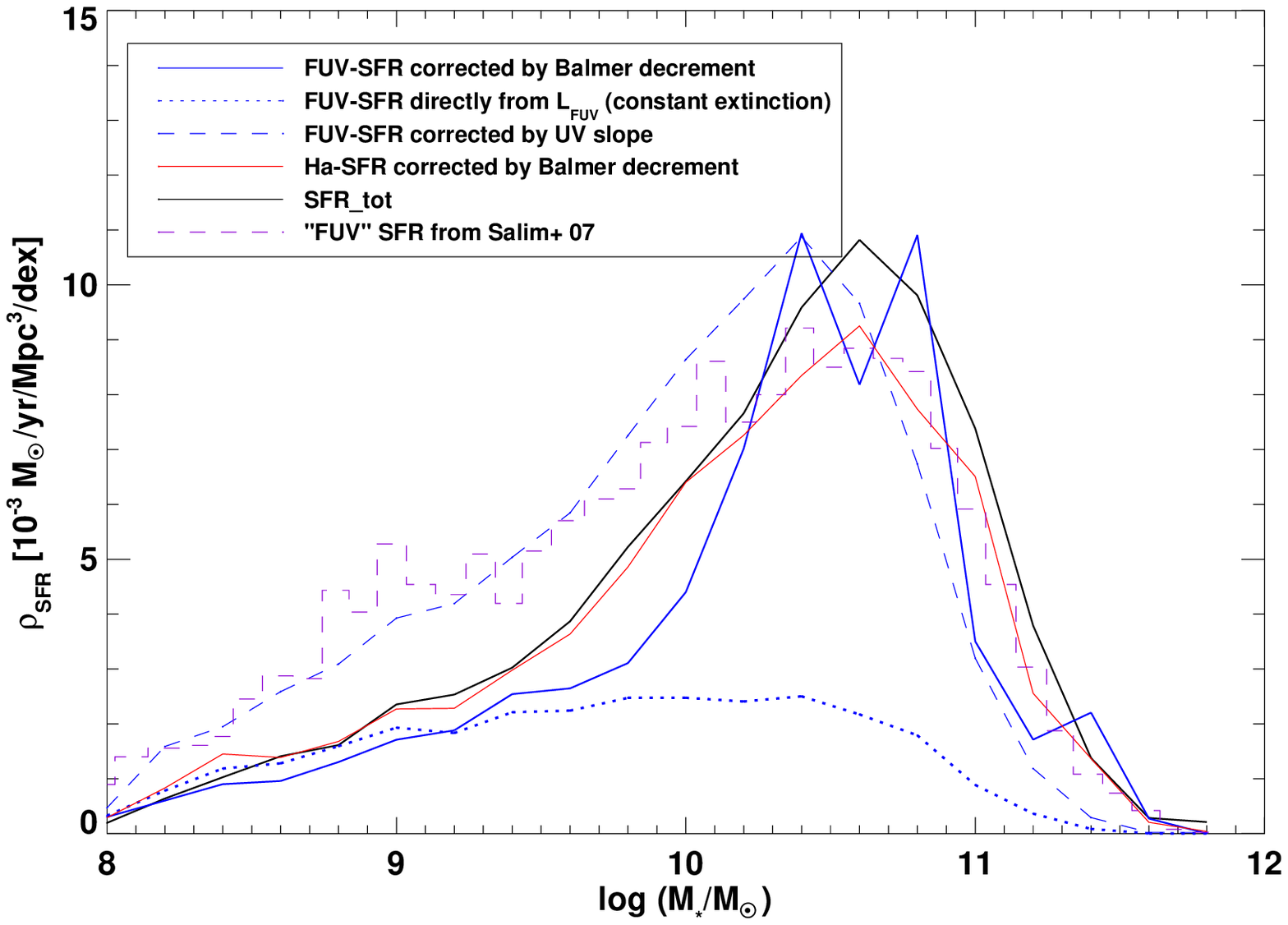}
\caption{As Fig.~\ref{fig:sfrd_lines}, but for the photometric SFR indicators.  Blue lines in the left panel show $u$-band derived SFRDs.  Solid line is corrected for extinction using the Balmer decrement, dotted line assumes constant extinction (equivalent to 1 mag at H$\alpha$), and dashed line assumes SFR-dependent extinction.  Right panel, blue lines show measurements from FUV.  Dashed line indicates extinction-corrected SFRD using the UV slope (NUV-FUV colour), dotted line assumes constant extinction (1 mag at H$\alpha$), and dashed line assumes SFR-dependent extinction.  Dashed histogram shows the (approximately FUV)-derived SFRD measured by \citep{Salim:2007rv} by SED fitting. 
\label{fig:sfrd_phot}
}
}
\end{figure*}

\subsection{Comparison of indicators}
\label{sec:comp_inds}

\subsubsection{Emission line indicators}
\label{sec:comp_em}

Fig.~\ref{fig:sfrd_lines} shows the comparison between the \ha~SFRD (left panel) measured using the Balmer decrement-correction, \ha$_{corr}$ (red solid line), the average mass-dependent extinction using our relation from Fig.~\ref{fig:avedust} (red dashed line) and a constant correction of $A_{H\alpha}=1$ mag (red dotted line). It is clear that the various different prescriptions (including simpler aperture corrections and using the same indicator for all galaxies regardless of spectral class) for \ha-inferred SFRs produce reasonable agreement with the modelling of \b04, particularly \ha$_{corr}$.  The biggest discrepancy is for the constant extinction method, as 1 mag of extinction at \ha~is clearly an overestimate for low mass (\ms$\sim$9) galaxies, and an underestimate for higher mass (\ms$\gsim$10), as can also be seen from Fig.~\ref{fig:avedust}.  The mass-dependent extinction correction is an improvement over the constant correction model, but overestimates the SFRD at masses above the peak.  We note that if we impose an upper limit of $A_{H\alpha}=2.0$ mag on the mass-dependent extinction, this overestimate of the SFRD at higher stellar masses disappears, and thus it may just be that the fit in Fig.~\ref{fig:avedust} is asymptoting to too high a value of the extinction.  The dashed green histogram shows the SFRD of \b04, measured from the larger area DR2. This is very slightly different from the result measured in the Stripe 82 area using their \sfrtot~indicator (black solid line).  The  peak of the former is offset to slightly lower masses (by $\approx$0.2 dex).  This may be due to evolution (see next section); the \b04~sample extending to slightly lower redshift; cosmic variance; or possibly differences in constructing the SFRD histogram. \b04~use the full probability distribution functions for the SFRs of each galaxy measured, whereas we have summed discrete values to construct the histogram.  Regardless of the cause, the difference between the Stripe 82 \sfrtot~SFRD and the \b04~result is small compared with the systematic offsets between indicators, which are of interest here.
 
The right panel of Fig.~\ref{fig:sfrd_lines} shows the comparison for the \oii-based SFRDs (blue lines). It is clear that the constant-extinction \oii-SFR (blue dotted line) is a very poor tracer of the SFRD.  The empirically-corrected \oii~SFRD traces the \ha$_{corr}$ and \sfrtot~SFRDs for masses up to \ms$\sim9.5$.  Thereafter it begins to systematically underestimate the \ha-based SFRDs, but still peaks at almost the same mass.  (In \S\ref{sec:sampsel}, we will show that this discrepancy is simply due to the \oii~flux limit, and the agreement between the \ha$_{corr}$~SFRD and that of the empirically-corrected \oii~SFRD is excellent.) In order to quantify the agreement between different indicators, we consider three measurements from the SFRDs: the position of the peak, the integrated SFRD within $8.5 \le $\ms$ \le 12.0$ and the integrated SFRD within $8.5 \le $\ms$ \le 10.4$ (since many indicators begin to diverge above this approximate position of the peak).  These values are listed in Table~\ref{table:sfrds}.  
 
Before discussing further the differences between the SFRDs using different estimators, it is pertinent to consider how the limiting SFR probed by each indicator affects the final SFRD in each mass bin.  i.e., whether the SFRD has converged.  Fig.~\ref{fig:cumsfrd} shows the cumulative SFRD as a function for \sfrtot~(solid black line) and the best estimators for each of \ha~(dashed red line) and \oii~(dot-dashed blue line).  The emission line estimators are plotted down to the limiting SFR, where this is defined as the median SFR which gives a S/N of 2.5 in the relevant emission line. In addition, the FUV SFR corrected using the UV slope is included (green dotted line), since it does not rely on any emission line estimator (nor on their associated aperture corrections, indeed not on any spectroscopic measurement except redshift) and is thus independent of the other methods.  

For the lowest stellar masses, all methods except the FUV give very similar answers.  As well as reaching comparable total SFRDs, each of these methods turn over at comparable SFRs ($\log(SFR)\sim-1$).  Above a mass of \ms$\sim$10, the estimated SFRDs all begin to converge to similar totals, however, towards increasing masses, it can be seen that the \oii~estimate is still rising as a function of limiting SFR, i.e., the \oii~SFRD does not converge.  This can be understood in terms of Fig.~\ref{fig:oii_ha_mass} (most clearly from the last panel).  At these stellar masses, the ratio between SFR (as traced by Balmer decrement-corrected \ha) to \oii~constant luminosity SFR is $\sim$10:1.  This means that \oii~is approximately an order of magnitude less sensitive to SFR than \ha~and becomes less so as stellar mass increases.  Thus, to the limit of the SDSS Stripe 82 data, the \oii-inferred SFRD has not converged to the \ha~value, for the highest mass galaxies.  For \ha, \sfrtot, and FUV, all estimators show turnovers in the SFRD at similar values of SFR, but the final totals achieved are different.  This implies that all values have converged, but to different limits reflecting systematic differences in the SFR estimators. These will be discussed in the context of the mass-dependent SFRD.
 
 \subsubsection{UV indicators}
\label{sec:comp_phot}

Fig.~\ref{fig:sfrd_phot} shows the same plots as Fig.~\ref{fig:sfrd_lines} for the photometric ($u$-band left panel and FUV right panel) SFR indicators.  To the methods already described in \S\ref{sec:emsfr} to measure photometric SFRs, we add the following methods to attempt to correct for dust extinction.  $A_u$, is approximated using the mass-dependent extinction relation (Eqn.~{\ref{eqn:massdep}) and a \citet{Calzetti:2000sp} extinction law; and also calculated using the Balmer decrement to calculate the extinction at \ha, as described above, and transforming this to 3270\AA~using the \citet{Calzetti:2000sp} law.  For the small subsample of  galaxies with insufficient S/N to measure the Balmer decrement, we estimate the extinction at \ha~from Eqn.~\ref{eqn:massdep}.  As a further possible improvement to the $u$-band SFR, we consider the SFR-dependent extinction correction.  We use the equations of \citet{Hopkins:2001db} to determine a relation between extinction and the observed UV luminosity.  The SFR was then determined using Eqn.~\ref{eqn:sfr_u}.   In addition, we use the Balmer decrement as described above to estimate the extinction at 1400\AA~and provide an alternate dust-corrected estimate of the FUV SFR.
 
The biggest discrepancy between indicators for low stellar masses occurs for the FUV SFRD compared with the other methods.  From Fig.~\ref{fig:sfrd_phot}, it can be seen that common to both sets of UV data, the estimators using only photometry (and photometric estimates for the dust correction) show an excess SFRD over the emission line based estimates at low stellar masses.  The UV slope-corrected FUV SFR (i.e. that plotted in Fig.~\ref{fig:cumsfrd}) uses the approximate calibration of dust extinction from the UV colour calibrated from the more detailed modelling of \citet{Salim:2007rv}.  Those authors showed that this approximation has a tendency to overestimate the SFR (compared with their SED modelling) at low values of SFR, and underestimate it at high values.  However, the dashed histogram shows the SFRD derived by \citet{Salim:2007rv} using their modelling, which is in very good agreement (certainly at the low mass end) with our estimate.  
Indeed, a galaxy-by-galaxy comparison with \ha$_{corr}$ (similar to Fig.~\ref{fig:oii_ha_mass}) shows excellent agreement between the FUV slope-corrected SFR and \ha$_{corr}$ with a small systematic underestimate of the latter by the former of $\sim$10\% but little mass dependence until \ms$\gsim10.5$, when the ratio drops further. This is better than might be expected for a galaxy-by-galaxy comparison, since the different timescales to which the UV and \ha~are sensitive would be expected to introduce significant scatter. 
Thus the result that the low mass end of the SFRD is predicted to be higher relative to the other methods when using UV data appears to be robust.  The $u$-band and FUV SFRDs using the Balmer decrement to estimate the dust are in closer agreement with the other estimators (which also largely or entirely use the Balmer decrement to estimate extinction).  Since our FUV SFRD agrees well with that of \citet{Salim:2007rv} and our \ha$_{corr}$ is close to that of \b04, we refer the interested reader to \citet{Salim:2007rv} for a discussion of the likely reasons for the discrepancies between these estimators.  They also found that the main disagreement is due to the way dust is estimated in the two methods.  They went through an extensive analysis trying to reconcile the dust estimators, and showed that the SED-fitting (i.e., UV-slope based) extinction estimate is noisier than that from the Balmer decrement, and thus likely overestimates the extinction at low stellar masses.  

\subsubsection{FIR indicators from the literature}
\label{sec:comp_fir}

At this point, it is interesting to compare the UV slope correction derived from \citet{Salim:2007rv}'s fitting to UV+optical SEDs with a similar relation derived from a different SFR indicator.  \citet{Cortese:2006rm} estimated UV extinction by comparing UV with total IR luminosities for a local sample of late-type galaxies with {\it GALEX} (UV) and {\it IRAS} (FIR) photometry. The combination of UV and FIR should sample both the unobscured and obscured components of star-formation, and so, when appropriately combined (e.g., \citealt{Martin:2007xe}, eqn.~7), should give the best estimate of total SFR for a galaxy. This ratio is usually denoted the Infrared excess (IRX $=  \log (L_{IR}/L_{UV})$. Taking the relation between IRX and the extinction in the {\it GALEX} FUV-band, $A_{FUV}$, from \citet{Buat:2005wd} and combining this with eqn.~4 \& 5 from \citet{Cortese:2006rm}, we obtain a relation between $A_{FUV}$ and (FUV-NUV) colour. This relation, based on IR estimates of the extinction, compares very well with the \citet{Salim:2007rv} relation, with an rms difference of 0.12 mag over the range of (FUV-NUV) colours of star-forming galaxies in our sample. We can approximate the relation as $A_{FUV} \approx 2.61(FUV-NUV)+0.409$ which may be compared with \citet{Salim:2007rv}'s eqn.~6 of $A_{FUV}=2.99 (FUV-NUV) + 0.27$ for galaxies passing their blue colour cut. Thus, our FUV estimate of the SFRD using the UV slope to estimate the extinction is relatively robust to the method by which the extinction is derived.

Therefore, using the Balmer decrement to correct the FUV SFR possibly represents the best combination of methods.  The FUV is a good tracer of star-formation (once corrected for extinction) and is much less affected by the presence of an AGN or the lack of a measurable \ha~emission line, and the Balmer decrement likely offers the best estimate of extinction.  We reiterate that for galaxies for which the lines necessary to measure the Balmer decrement possess insufficient S/N, we use the average mass-dependent extinction, and thus are not limited to requiring \ha~such as in the \ha-selected sample.  The fact that this hybrid method follows quite closely \ha$_{corr}$ reinforces the suitability of this method. 

A further point worth noting from Fig.~\ref{fig:sfrd_phot} is that the $u$-band SFRD, even uncorrected for dust, provides a surprisingly accurate estimate of the SFRD compared with the more observationally expensive estimators.  A mass-dependent extinction correction such as Fig.~\ref{fig:avedust} (under the assumption that the locally derived value is applicable) converted to the appropriate wavelength would provide an even better estimate, and thus these are useful indicators for higher redshift surveys.  A better estimate still can be made using bluer data (FUV) plus a correction for the UV slope as described in \citet{Salim:2007rv}.  The greater sensitivity of the bluer data to the dust correction can clearly be seen by comparing the constant luminosity estimators (dotted lines) in both panels of Fig.~\ref{fig:sfrd_phot}.

\begin{table}
\caption{Properties of the SFRD for the various different SFR indicators.  Columns are: method used; \ms~at which the SFRD peaks; integrated SFRD over the widest  mass range considered ($8.5 < $\ms$ < 12.0$); integrated SFRD from the completeness limit to the approximate peak ($8.5 < $\ms$ < 10.4$) -- this latter range is chosen since the largest disagreement occurs at mass higher than the peak SFRD.  Integrated SFRDs are given in units of $\times10^{-3}M_\odot yr^{-1}Mpc^{-3}$ .  ($^*$ - histogram is bimodal, this value is midway between the two peaks. $^{**}$ - exact normalisation depends on dust conversion from \ha~to UV, see \S\ref{sec:dustcorr})}
\label{table:sfrds}
\begin{tabular}{lrrr}
\hline
Method & peak mass & $\int_{8.5}^{12}\rho dM$ & $\int_{8.5}^{10.4}\rho dM$ \\
\hline
              \sfrtot  &  10.6  & 15.24  &  5.86  \\
    \ha~Balmer decr.  &  10.6  & 13.41  &  5.65  \\
                 \ha (const.) &  10.0  & 12.94  &  7.79  \\
       \ha~mass dep.  &  10.4  & 16.11  &  7.92  \\
   \oii$_{emp,corr}$  &  10.4  & 10.58  &  5.18  \\
 \oii$_{MB,corr}$  &  10.0  &  7.19  &  4.90  \\
                \oii  (const.) &   9.8  & 13.25  & 10.61  \\
    $u$ Balmer decr.$^{**}$  &  10.8  & 15.16  &  7.06  \\
       $u$ mass dep.$^{**}$  &  10.4  & 18.53  &  9.49  \\
        $u$ SFR-dep.  &  10.4  & 15.45  &  8.23  \\
                 $u$  (const.)$^{**}$ &  10.0  & 13.36  &  8.15  \\
        FUV UV slope  &  10.4  & 16.21  &  8.78  \\
            FUV Balmer decr.$^{**}$  &  10.4$^*$  & 13.00  &  4.24  \\
          FUV const.$^{**}$  &  10.4  &  9.49  &  6.10  \\
                \b04  &  10.75  & 14.78  &  6.53  \\
\hline
\end{tabular}
\end{table}

\subsubsection{Other empirical corrections}
\label{sec:othemp}

\citet{Moustakas:2006wp} propose an empirical  correction from \oii~luminosity to SFR using the rest-frame $B$-band luminosity, $L_B$.  They give a method for computing rest frame $B$-band luminosities that can be calculated from SDSS $g$ and $r$ photometry (their eqn.~1 and {\sc kcorrect}) and plot the ratio between \ha-derived SFR and $L_{\rm [OII]}$ as a function of $L_B$ (their fig.~19).  We construct $L_B$-corrected \oii~SFRs for the Stripe 82 data following this approach and interpolating their fig.~19 to derive a correction for each galaxy on the basis of its $L_B$.  This produces the dotted curve in Fig.~\ref{fig:sfrd_lines}.  It can be seen that this correction severely underestimates the SFRD at high stellar masses.  

\citet{Weiner:2007tf} derive a correction for \oii~SFR based on rest-frame $H$-band magnitude, from observed $K$-band data at z$\sim$1, by comparing the observed \oii~luminosity with SFR derived from \ha~emission.  They give this correction as 0.23 dex mag$^{-1}$ of rest-frame $H$-band.  We synthesise rest-frame $H$-band magnitudes from the stellar masses and rest-frame $(g-r)$ colours described in \S\ref{sec:mass}, using table 7 of \citet{Bell:2004lb}.  We set the normalisation of the $H$-corrected \oii~SFR by requiring the SFRD at \ms$\sim$10 to agree with that of \oii$_{corr}$.  This gives a curve (not plotted for clarity) which matches very closely over the whole mass range considered to that of the \citet{Moustakas:2006wp} correction plotted in Fig.~\ref{fig:sfrd_lines}, and so these two indicators may be considered equivalent.  Neither of these corrections adequately capture the large corrections necessary to \oii~SFR at high stellar masses.  

Recently \citet{Pannella:2009uj} estimated the median SFR in bins of stellar mass at z$\sim$2 using photometric redshifts to select galaxies and determine stellar mass, and stacked 1.4GHz radio emission to determine total SFR in these mass bins.  They found that the SFR is almost a linear function of \ms, i.e. a power law in $M_*$ with an exponent of order unity.\footnote{\citet{Dunne:2009bg} came to similar conclusions using a similar technique on a different field, but with a somewhat shallower relation.}  By comparing the SFR estimated from the rest-frame 1500\AA~luminosity with the stacked radio SFR (assuming the latter is an accurate measure of the total SFR), they derived an empirical relation for the extinction as a function of stellar mass (assuming the difference between the two SFRs is entirely due to extinction in the 1500\AA~flux estimates):

\begin{equation}
A_{1500} = 4.07 \log(M_*/M_\odot) -39.32.
\label{eqn:pannella}
\end{equation}
Applying this (rather steep) correction as a function of stellar mass (converted using our assumed extinction described in \S\ref{sec:dustcorr}) to our Stripe 82 $u$-band estimated SFR results in a vast overproduction of the z$\sim$0.1 SFRD. The attenuation predicted at \ms$\sim$11 is over 6 magnitudes, or  a factor of 250, and thus the peak of the $u$-band SFRD would be off the top of the plot in Fig.~\ref{fig:sfrd_phot}.   Similarly, using our extinction law to convert the 1500\AA~continuum attenuation to attenuation in the gas at the wavelength of \oii~or \ha~would lead to large overestimates in the SFRD for these other indicators.  Clearly, the \citet{Pannella:2009uj} prescription is not a good estimate for the mass-dependent extinction correction in the local Universe.  This suggests that either the dust properties as a function of mass have evolved strongly between z$\sim$2 and z$\sim$0, such that extinction is much lower now; or possibly that observational systematic errors in the z$\sim$2 data lead to an overestimate of the correction.  If the stacked galaxy samples contain contamination in the radio data (such as might be produced by low luminosity AGN) then the radio-SFR (and hence extinction correction) would be overestimated (see \citealt{Pannella:2009uj}).

We note that other groups \citep{Martin:2007xe,Buat:2007zr,Iglesias-Paramo:2007db} have used the infrared excess, IRX, as a measure of extinction and found more modest UV extinctions than predicted by eqn.~\ref{eqn:pannella}. These works do not explicitly give a comparable relation, fitted as a function of stellar mass, but we can use \citet{Martin:2007xe}'s IRX versus stellar mass data and convert IRX to $A($\ha$)$ as follows. We first take the relation between IRX and UV slope, $\beta$ from \citet[][eqn.~5]{Cortese:2006rm}.  Now, we just need to convert between UV slope, $\beta$ and $A($\ha$)$. \citet{Cortese:2006rm} present two relations between these quantities, one fitted from their own data (their eqn.~9) and the other taken from \citet{Calzetti:2000sp} (the former's eqn.~8). However, we may fit our own relation between $\beta$ (simply related to the (FUV-NUV) colour, e.g., \citealt[][eqn.~4]{Cortese:2006rm}) and $A($\ha$)$ taken from the Balmer decrement. Doing this we find $\beta \approx 1.26A(H_{\alpha}) -1.79$ but the scatter is large and the fit is not well-constrained. We show the results of taking the z$=$0.1 stellar mass--IRX data (adjusted to our IMF) from fig.~7 of \citet{Martin:2007xe} and converting IRX to $A($\ha$)$, as just described, as the large, filled circles in Fig.~\ref{fig:avedust}. As can be seen, the points are in reasonable agreement with our estimates of the extinction as a function of stellar mass derived directly from \ha. If instead of using the relation we measure in Stripe 82 we take the relation from \citet{Calzetti:2000sp} ($\beta = 0.75A({\mathrm H}\alpha) -1.80$), the points would be in even better agreement with the solid line of Fig.~\ref{fig:avedust}. Using the corresponding relation fitted by \citet{Cortese:2006rm} would predict much lower \ha~extinction at a given stellar mass, and be in considerably worse agreement than the points shown. This may suggest that the subsample of the \citet{Cortese:2006rm} survey possessing optical spectroscopy (to allow \ha, H$\beta$ to be measured) might not be representative of the larger surveys of \citet{Martin:2007xe} (COMBO-17) and/or the present work (Stripe 82). However, selection bias could be present at any stage of the conversions we have performed (the different surveys use some combination of UV-selection, morphological selection, mass selection, etc.). A detailed comparison of the IR-derived extinction is beyond the scope of this paper. Indeed, ideally one would like to have FIR data for exactly the same galaxies as all the other indicators (i.e., Stripe 82 galaxies) rather than resorting to relations derived from other datasets. Unfortunately, wide-field FIR imaging does not exist for Stripe 82 and so we cannot make a direct comparison. 

\subsection{Possible systematic errors}

\subsubsection{Sample selection}
\label{sec:sampsel}

The main source of differences between the \sfrtot-, \ha-, and \oii-based SFRDs is the selection criterion.  Each of the latter two measurements requires a S/N$>$2.5 in the corresponding line in order to be considered in the sample, whereas the \sfrtot~measurement is based on two different techniques (described in \S\ref{sec:sfrtot}), switching to the alternate technique when the emission line flux drops below the minimum requirement, or is otherwise corrupted.  This is illustrated in Fig.~\ref{fig:sfrd_thresh}.  The solid lines again show the SFRDs based on \sfrtot, \ha, and \oii, as in previous figures (black, red and blue lines, respectively). The dashed black line shows the SFRD based on \sfrtot, but now with the \ha~sample's selection criterion (i.e., S/N(\ha)$>$2.5).  It can be seen that this agrees very well with the SFRD from \ha, implying that the main difference between the two samples is the minimum significance in the \ha~line.  The largest difference occurs around the peak of the SFRD, for galaxies \ms$\sim$10.5.  This is because stellar absorption is strongest for the most massive (brightest) galaxies, and thus the lowest SFRs in \ha~at these masses are hidden by underlying absorption in the stellar continuum.  Thus, it may be better to think in terms of an \ha~equivalent width limit as well as a S/N limit.  For these galaxies with weak SFRs and strong Balmer absorption (and therefore S/N$<$2.5), \sfrtot~switches to using $SFR_d$ instead of $SFR_e$.  From the comparison in the previous section we (and also \citealt{Salim:2007rv}) have argued that \sfrtot~likely overestimates the SFR when $SFR_d$ is used instead of $SFR_e$ (i.e., for the ``no-\ha" or AGN component objects).  The difference between the solid red and black dashed lines now show directly what fraction of the SFRD use this $SFR_d$ estimate.  

Similarly, the dashed red line shows the \ha~SFRD with the requirement that S/N(\oii)$>$2.5.  This brings the SFRD much closer to that measured by the \oii-selected sample, albeit somewhat lower at the high mass end.  This means that the empirically-corrected \oii~SFRD shown in Fig.~\ref{fig:sfrd_lines} will provide an excellent proxy for the \ha$_{corr}$ SFRD, provided the \oii~flux limit is sufficiently deep.

\begin{figure}
{\centering
\includegraphics[width=85mm]{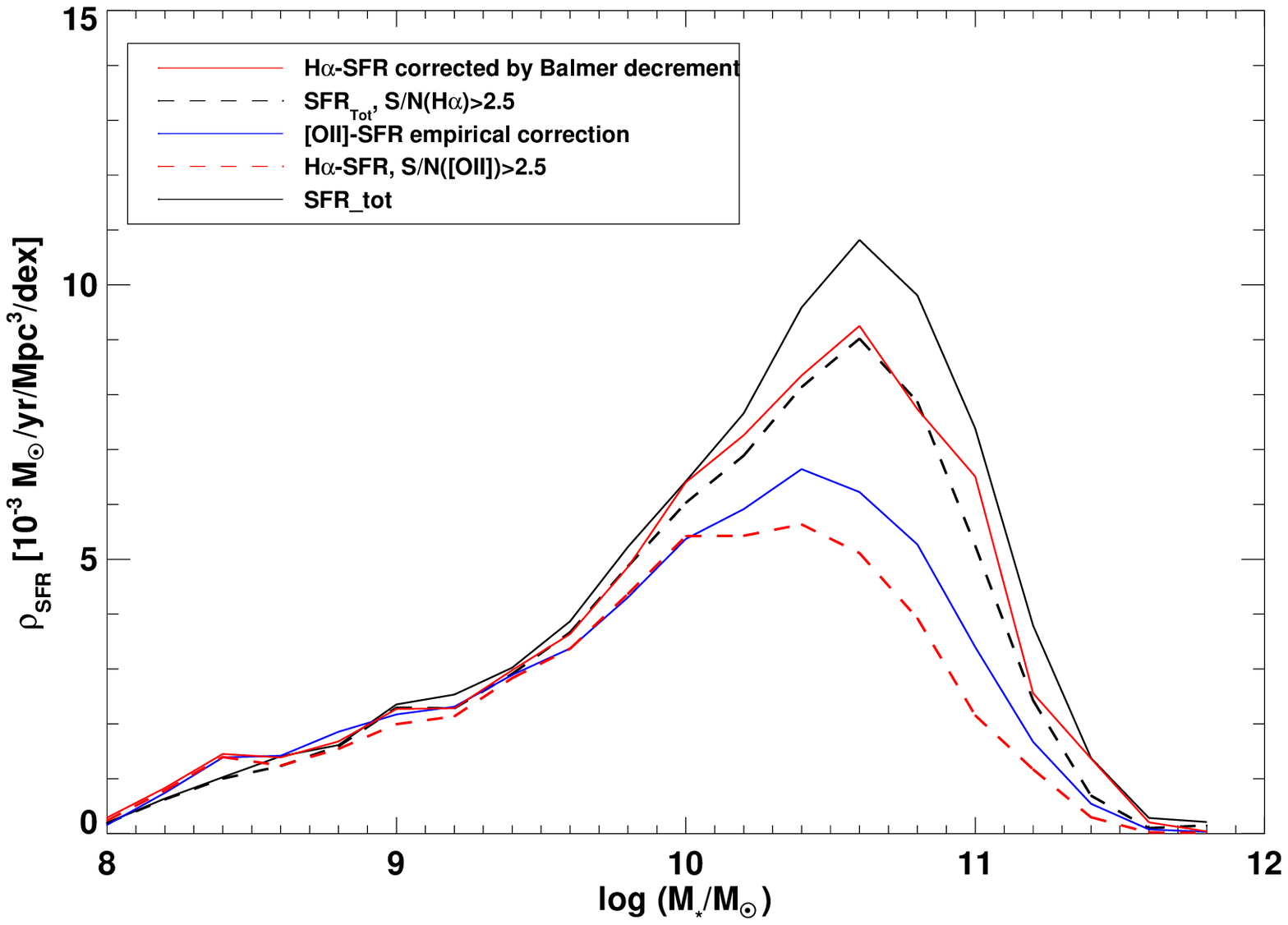}
\caption{As Fig.~\ref{fig:sfrd_lines}, but imposing different S/N cuts on the sample selected.  The solid black, red and blue lines show the nominal \sfrtot, \ha$_{corr}$, and \oii$_{emp,corr}$ SFRDs, as in Fig.~\ref{fig:sfrd_lines}.  The dashed black line shows the SFRD measured from \sfrtot, but now imposing the \ha~selected S/N cut.  This can be seen to agree well with the \ha-selected SFRD.  The red dashed line shows the SFRD measured from \ha, but imposing the \oii~selection (S/N(\oii)$>$2.5).  This agrees quite closely with the \oii-derived SFRD, but the latter gives a slightly higher SFRD toward higher masses.  See text for discussion.
\label{fig:sfrd_thresh}
}
}
\end{figure}

\subsubsection{Completeness, evolution and aperture correction}
We discuss in detail the effects of survey completeness and evolution in Appendix \ref{sec:completeness} and aperture corrections in Appendix \ref{sec:apcorr}.  For the present discussion, we note that the fact that our SFRD estimate using \sfrtot~in Stripe 82 is in good agreement with the SFRD measurement of \b04~from the DR2 area means that our completeness and method of aperture correction give comparable results and thus our corrections cannot be egregious.  Furthermore, our results presented have not been corrected for evolution and our non-evolution corrected, $0.032<z<0.20$ sample agrees with the evolution-corrected $0.005<z<0.22$ \b04~sample, implying that evolution within our sample is a small effect.

\subsubsection{AGN contamination}
\label{sec:class}

The effect of AGN contamination on the total SFRD inferred from emission lines such as \oii~and \ha~is complicated to assess.  \b04~adopted the approach of not using emission line measurements to measure SFR where there may be significant contribution to the line flux from AGN light, as determined from line ratios using the BPT diagram; and instead used a relation derived from star-forming galaxies.  Furthermore, \b04~use this relation for any galaxy not classed as star-forming (inlcuding low S/N star-forming and ``no \ha") and apply the relation between D4000 and SFR/stellar mass calibrated from star-forming galaxies.  \citet{Salim:2007rv} have suggested, based on their UV-derived SFRs, that this relation used by \b04~may not be directly applicable to non-star-forming galaxies (see in particular their fig.~3).  Since in the current work we are primarily interested in applying simple SFR calibrations to high redshift, we adopt the following.  At high redshift, the common practice is to use whatever data is available to identify galaxies in the sample which may contain an AGN (e.g., MIR colours, X-ray emission; typically a detailed decomposition of the contribution from an AGN to the line flux using, for example the BPT diagram, is not possible) and reject them from the analysis.  Thus, we use the line ratio classification from the current sample to reject those objects residing in the `AGN' region of the BPT diagram and measure the SFRD in the same way as before.  The best \oii~and \ha~measurements are shown for this class of galaxies in the left panel of Fig.~\ref{fig:sfrd_agn}.  Comparing this with Fig.~\ref{fig:sfrd_lines} shows that excluding AGN makes very little difference to the total SFRD.  Conversely, including AGN in the SFRD measurement and naively converting their line luminosities to SFR does not lead to a significant overestimate of the SFRD.  If, however, only the star-forming class is considered, the total SFRD now drops appreciably.  This is shown in the right panel of Fig.~\ref{fig:sfrd_agn}.  It clearly shows that a more reasonable estimate of the total SFRD may be obtained by either taking all galaxies (Fig.~\ref{fig:sfrd_lines}) or all but those whose light is dominated by AGN emission (Fig.~\ref{fig:sfrd_agn}, left panel) and applying the same SFR estimator to the resulting sample, rather than omitting all objects with any hint of an AGN contribution (Fig.~\ref{fig:sfrd_agn}, right panel).  

As an independent estimate of the effect of AGN, the photometric SFR indicators ($u$-band and FUV) use colour selection to decide whether a galaxy is star-forming or not, independent of whether its emission lines possess sufficient S/N to assign a spectral class.  AGN and composite (star-forming$+$AGN) classes lie predominantly in the ``green valley" between the red and blue sequences in UV-optical CMDs or on the red-sequence in optical CMDs (e.g., \citealt{Salim:2007rv}) and thus our blue sequence selections will efficiently reject AGN.  More importantly, due to the steep power law nature ($\propto\nu^{-1.5}$) of non-thermal emission from AGN (compared with the more Blackbody-like nature of the stellar light), the contamination to the rest-frame UV flux from AGN is much less problematic than the contamination to emission lines such as \ha~and \oii.

\begin{figure}
{\centering
\includegraphics[width=90mm]{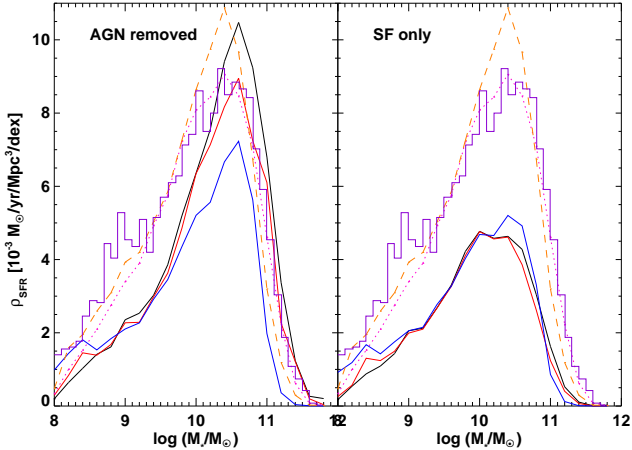}
\caption{SFRD as a function of stellar mass for different classes of galaxies.  Solid histogram shows \sfrtot~estimate, red and blue curves show best estimates (Balmer decrement corrected and best correction) for \ha~and \oii~SFRs respectively. Left panel shows these measures with the AGN class (based on line ratios) removed, right panel shows the same for only those galaxies in the `star-forming' class (after removing AGN, composite and low S/N classes).  For both panels, the same photometric measures of SFRD (orange dashed line shows slope-corrected FUV; magenta dotted lines shows Balmer decrement-corrected $u$-band; and purple histogram shows result from \citealt{Salim:2007rv}) are shown as in Fig.~\ref{fig:sfrd_phot}.  The photometric samples are based only on galaxy colour to determine which galaxies are star-forming, and do not use the line ratio classification.  See text for details.  
\label{fig:sfrd_agn}
}
}
\end{figure}

\subsubsection{Obscured star-formation}

The main disadvantage of all the SFR indicators considered in this work is that a correction must be made for the effects of dust attenuation and that none of the indicators is directly sensitive to dust-obscured star-formation.  In order to directly trace dust-obscured star-formation, a dust-insensitive  indicator such as radio continuum emission \citep{Condon:1992gd} or Mid- or Far-IR emission (e.g., \citealt{Kennicutt:1998pa}) should be employed.  Unfortunately, the Stripe 82 region does not possess sufficient overlapping radio or MIR/FIR data to allow a direct test of the dust-corrections used.  However, \citet{Hopkins:2003gs} studied a sample of 106\,000 galaxies in the SDSS Early Data Release overlapping with the FIRST 1.4GHz radio survey.  They found that the Balmer decrement-corrected \ha~SFR gave reasonable agreement with the radio-derived SFR, although the fact that only 3079 galaxies from this large sample had measurable radio luminosities (of which 791 were due to star-formation rather than AGN activity) shows the sensitivity limitation of using radio emission as a tracer of star-formation.  

In \S\ref{sec:comp_inds}, we have shown that estimates of extinction using FIR results scaled from other surveys give comparable results to the optical/UV indicators we employ in this work. Obviously overlapping extinction-insensitive indicators such as FIR would be a desirable addition to the Stripe 82 data to allow a direct comparison of these measures.

\subsection{Application to higher redshift samples}
\label{sec:highz}

One of the main motivations for this work is to be able to use the corrections derived in the local Universe to correct the mass-dependence of the SFRD at higher redshift.  At z$\sim1$, several surveys are now starting to produce statistical samples of galaxies with spectroscopic redshifts, stellar masses and \oii-derived SFRs.  The compilation presented recently in \citet{Davies:2009nx} used three surveys to construct a sample spanning 8$\lsim$ \ms $\lsim$12 at z$\sim$1 (ROLES, GDDS, DEEP2, respectively \citealt{Davies:2009nx,Juneau:2005ft,Conselice:2007ns}).  The first two surveys, ROLES and GDDS, use only \oii~to measure SFR, whereas DEEP2 uses a combination of \oii~and 24$\mu$m.  For the moment, we just consider ROLES and GDDS' \oii-derived SFR which use directly Eqn.~\ref{eqn:sfr_oii} (i.e., constant luminosity conversion), but with a small correction for IMF.  For this section only we switch to our preferred IMF of \citet{Baldry:2003ue} (BG03).  \oii~SFRs can be transformed from the Kroupa to BG03 IMF using $\log SFR_{BG03} = \log SFR_{Kroupa}-0.187$ and stellar masses using $\log M_{*,BG03} = \log M_{*,Kroupa} -0.08$.  This choice of IMF also means the \citet{Davies:2009nx} numbers may be used directly, and we reproduce the results from their fig.~2 in Fig.~\ref{fig:sfrdz1}.  The raw measurements of \oii~SFRD using the constant extinction prescription are shown as open circles with error bars.  As shown in \S\ref{sec:empoii}, the empirical mass-dependent correction Eqn.~\ref{eqn:oiipoly} provides an improved estimate of the \oii-inferred SFRD over the simplistic assumption of constant dust extinction.  This correction (solid symbols) {\it lowers} the SFRD in low mass galaxies (\ms $\lsim$10) and raises the value in higher mass galaxies.  As such, the conclusion of \citet{Davies:2009nx} that a turnover of the z$\sim$1 SFRD occurs below \ms$\sim$10 is strengthened.  

Under the assumption that the SFR function has the same shape at z$\sim$1 as z$\sim$0, we can use Fig.~\ref{fig:cumsfrd} to ask if the SFRD has converged in the high redshift sample.  In SDSS, a limiting SFR of $\sim$1 M$_\odot~yr^{-1}$ is sufficient at \ms$\sim$10.  Using our empirical correction (Eqns.~\ref{eqn:oiipoly} and \ref{eqn:sfr_oii}) applied to GDDS' conservative \oii~flux limit \citep{Juneau:2005ft} yields a limiting SFR of around 2 M$_\odot~yr^{-1}$.  Similarly, ROLES would have a corrected limit of $\sim$0.2 M$_\odot~yr^{-1}$ at \ms$\sim$9.  Fig.~\ref{fig:cumsfrd} shows that a limiting SFR of $\sim$0.1 is sufficient for the SFRD to have converged locally.  So, both z$\sim$1 surveys are quite close to this local limit.  However, the SFRD is on average $\gsim$3 times higher at z$\sim$1 than z$\sim$0.  Thus, if the shape of the SFR function is the same at both epochs and the required SFR limits are higher by a similar amount, these higher redshift surveys are comfortably above the SFR limit needed for the SFRD to have converged.    

We repeat that although we have used stellar masses from SDSS derived in a very different way from that commonly done at high redshift, if we switch to SED-fitted masses (indeed, the same code as used to fit the ROLES and GDDS data), our results are unchanged.  

It is prescient to ask if the empirical correction (Eqn.~\ref{eqn:oiipoly}) derived at z$\sim$0 is applicable to galaxies at z$\sim$1.  The correction is likely due to the metallicity dependence of \oii, the mass-dependence of the \oii/\ha~ratio and dust extinction.  Recent observations of the mass--metallicity relation at z$\sim$1 (e.g., \citealt{Savaglio:2005hp}, \citealt{Cowie:2008ob}) suggest that the relation is consistent with only mild evolution from that measured at lower redshifts, and so, since the metallicity correction is detemined via this relation, the metallicity dependence might be expected to be similar.  Unfortunately, data to make a stringent test of all these factors does not currently exist. However, we can make an approximate check by taking the SFRD from DEEP2 (open diamonds with error bars) using 24$\mu$m SFRs \citep{Conselice:2007ns} and comparing this with the empirically-corrected \oii~SFRs.  Since the 24$\mu$m emission probes radiation from star formation reprocessed via dust emission, the combination of \oii$+$24$\mu$m (as used in \citealt{Conselice:2007ns}) should be a good proxy for total SFR.\footnote{In this case, \citet{Conselice:2007ns} use \oii~uncorrected for extinction effects, since the obscured star-formation is directly probed by the 24$\mu$m emission.  Since the bulk of our empirical correction accounts for dust, it is appropriate to use the \citet{Conselice:2007ns} SFR without our empirical correction to \oii.}  These can be seen to appear consistent with the corrected \oii~measurements in Fig.~\ref{fig:sfrdz1}.  Although the mass range of the DEEP2 data is limited to the very highest mass galaxies, this is where the correction should be largest and thus the agreement, albeit within the broad uncertainties, is reassuring.  A more thorough test of our correction at higher redshift will require, for example, overlapping \ha~and \oii-SFRs for a sample of galaxies spanning a range of stellar masses and such work is currently underway.

\begin{figure*}
{\centering
\includegraphics[width=140mm]{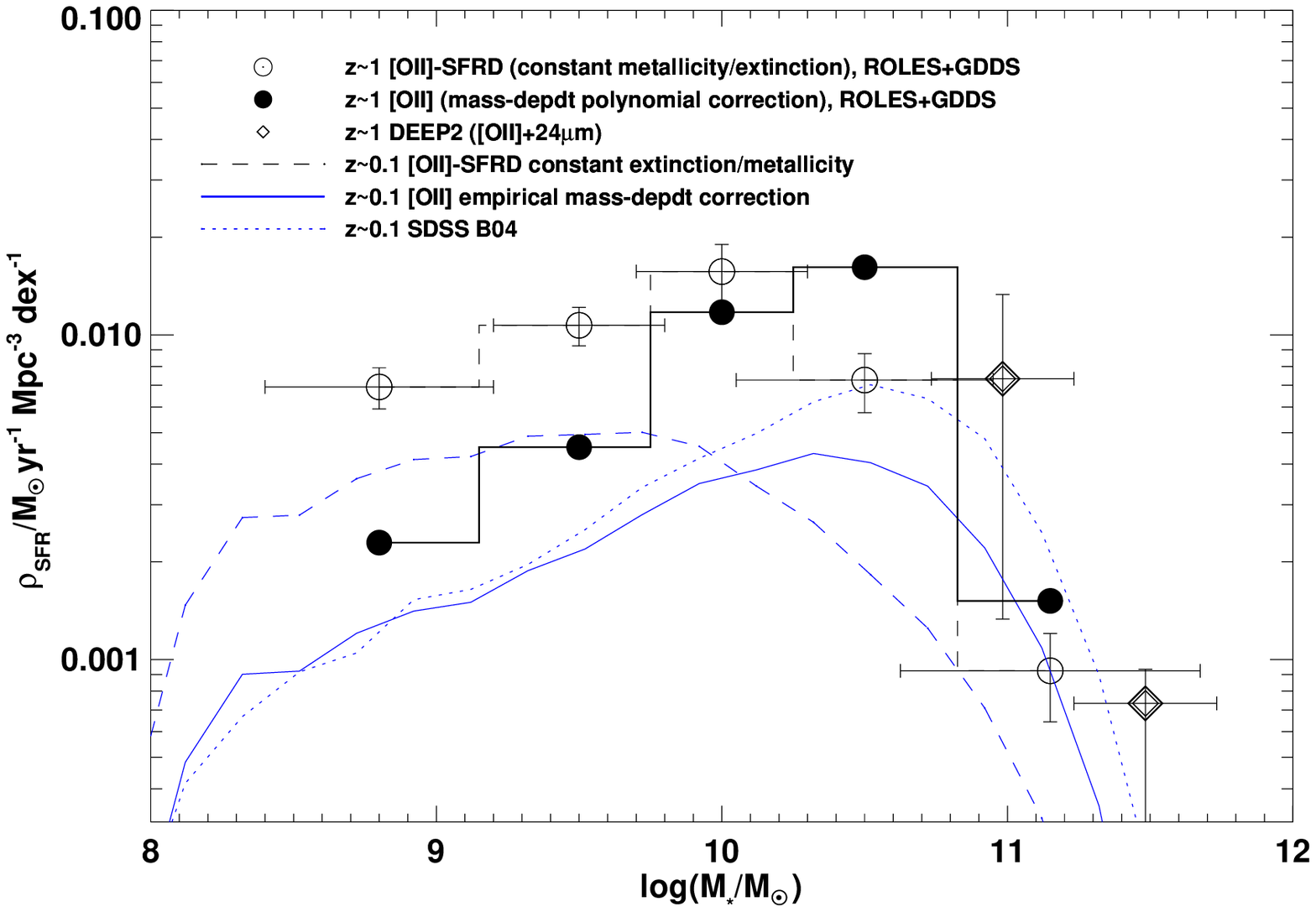}
\caption{SFRD versus stellar mass for the z$\sim$1 samples from \citet{Davies:2009nx} (ROLES, \ms$<$10.2) and \citep{Juneau:2005ft} (GDDS, \ms$>10.2$).  Open circles with error bars show the \oii-inferred SFRD from these surveys using the constant luminosity conversion \oii~SFRD.  Filled circles show the same measurements corrected using our empirical correction from Eqn.~\ref{eqn:oiipoly}.  Open diamonds with error bars show the \oii+24$\mu$m SFRD from DEEP2 \citep{Conselice:2007ns}.  The curves show the \oii-derived local SFRD from SDSS Stripe 82 using constant luminosity conversion (dashed line) and \oii$_{corr}$ using our empirical correction (solid line). The dotted line shows the \b04 estimate for reference.
\label{fig:sfrdz1}
}
}
\end{figure*}

\section{Conclusions}

We have compared SFR indicators based on \ha, \oii, $u$-band and FUV luminosity using the SFRD as a function of stellar mass for $\sim$50\,000 in the SDSS Stripe 82 region ($0.032<z<0.20$).

Our main conclusions are:

$\bullet$ Using Balmer decrement-corrected \ha~luminosity and a simple $u$-band based aperture correction for all galaxies irrespective of class yields a SFRD in good agreement with the detailed modelling of \b04.  

$\bullet$ In the absence of Balmer decrement information, a simple mass-dependent prescription (Eqn.~\ref{eqn:massdep}) can be used for correcting SFR estimators for extinction.

$\bullet$ \oii~luminosity works well as a proxy for \ha~luminosity, in a statistical sense, but previous empirical corrections underestimate the correction required toward high stellar mass (\ms$\gsim10$).  We present a new empirical correction as a function of stellar mass which is calculated on a galaxy-by-galaxy basis for spectroscopically classified star-forming galaxies, and is thus insensitive to uncertainties in aperture correction, AGN contamination, etc.:

\[
SFR_{emp,corr}/(M_\odot yr^{-1}) = \frac{L([{\rm O\,II}])/(3.80 \times 10^{40} erg\,s^{-1})}{a\,\tanh[(x-b)/c] +d},\]
where $a=-1.424$, $b=9.827$, $c=0.572$, and $d=1.700$, for a Kroupa IMF.

$\bullet$ Our empirical correction to \oii~is shown to reconcile the \ha~and \oii~SFRDs as a function of stellar mass.   After applying this correction, the \oii~SFRD is still somewhat lower ($\sim30\%$) than other methods, such as \ha, but the primary reason for this difference is the decreased sensitivity of \oii~to a given SFR threshold (set by the flux limit of the SDSS spectroscopy) at the high stellar mass end.

$\bullet$ Photometric ($u$ and FUV) SFRDs (for galaxies with spectroscopic redshifts) give reasonable agreement with the emission line based SFRDs, but the effects of extinction toward high stellar masses become more important, particularly for the FUV.  A reasonable estimate of the extinction can be made from the UV slope. All the photometric indicators predict a greater SFRD in low mass galaxies than the emission line indicators.  The $u$-band luminosity, even assuming the simple constant-extinction model, is a powerful SFR proxy.  

$\bullet$ We have estimated the limiting star-formation rate threshold required for the SFRD to converge and demonstrated how this may be applied to higher redshift surveys, assuming that the SFR function in a given mass bin has the same shape as that in the local Universe.

$\bullet$ Applying our empirical mass-dependent correction for \oii~SFR to an \oii-selected sample at z$\sim$1 shows that the turnover in the SFRD of low mass (\ms$\lsim$9) galaxies seen by \citet{Davies:2009nx} is in fact even more significant than reported in that paper, once the effects of metallicity and extinction on \oii~luminosity are taken into account. 

Mass dependent systematic effects (such as extinction) are of vital importance to studies of galaxy evolution.  Neglecting such effects may lead to incorrect trends as a function of stellar mass.  Using corrections derived locally, such as those presented here, is an improvement over current methods which assume, for example, a constant conversion between \oii~luminosity and SFR.  \oii-selected surveys can potentially provide a powerful and efficient probe of star-formation in the z$\gsim$1 Universe (e.g., \citealt{Gilbank:2010hc}).  Verifying the accuracy of these empirical calibrations will require the cross-comparison of different SFR indicators at higher redshifts.  In particular, the inclusion of \ha, \hb, \oiii~and [N\,{\sc ii}]~data using 8-m class, NIR, multi-object spectroscopy on statistical samples will be a critical test.

\section*{Acknowledgments}

We thank the referee for thorough and thoughtful comments which improved the content and presentation of this paper. We thank Jarle Brinchmann for useful discussion on the details of the \citet{Charlot:2001gj} models, and Chris Simpson and Phil James for helpful suggestions.

This research was supported by an Early Researcher Award from the
province of Ontario, and by an NSERC Discovery grant.

Funding for the SDSS and SDSS-II has been provided by the Alfred P. Sloan Foundation, the Participating Institutions, the National Science Foundation, the U.S. Department of Energy, the National Aeronautics and Space Administration, the Japanese Monbukagakusho, the Max Planck Society, and the Higher Education Funding Council for England. The SDSS Web Site is http://www.sdss.org/.

The SDSS is managed by the Astrophysical Research Consortium for the Participating Institutions. The Participating Institutions are the American Museum of Natural History, Astrophysical Institute Potsdam, University of Basel, University of Cambridge, Case Western Reserve University, University of Chicago, Drexel University, Fermilab, the Institute for Advanced Study, the Japan Participation Group, Johns Hopkins University, the Joint Institute for Nuclear Astrophysics, the Kavli Institute for Particle Astrophysics and Cosmology, the Korean Scientist Group, the Chinese Academy of Sciences (LAMOST), Los Alamos National Laboratory, the Max-Planck-Institute for Astronomy (MPIA), the Max-Planck-Institute for Astrophysics (MPA), New Mexico State University, Ohio State University, University of Pittsburgh, University of Portsmouth, Princeton University, the United States Naval Observatory, and the University of Washington.

\appendix
\section{aperture corrections}
\label{sec:apcorr}

As discussed in \S\ref{sec:linefluxes}, it is necessary to extrapolate from flux measured within the fibre to total flux.  The total SFRs measured from emission lines can be dominated by the flux outside the fibre, particularly for the most massive (and hence largest) galaxies.  We have adopted a simple scaling based solely on the ratio of the $u$-band Petrosian flux (which should be a good approximation to the total flux) to the $u$-band flux measured within the fibre.  We choose the $u$-band since this is likely to give the best proxy for SFR.  This scheme is similar to that adopted by \citet{Hopkins:2003gs}, except that they chose to use a $u$-band scaling only for \oii~and $r$-band for \ha; whereas we use a $u$-band scaling for both measurements.  Experimentation with other bands showed that using $g$ or $r$ band to aperture correct gave comparable results to using $u$; the biggest difference was that the $g$/$r$ aperture correction becomes smaller than the $u$-band correction by $\sim$10-20\% for high stellar mass (\ms$\gsim$10.5) objects.  

Fig.~\ref{fig:apcorr} shows the aperture correction ($u$ Petrosian flux/$u$ fibre flux) as function of redshift and mass.  Coloured points show galaxies with stellar masses in the three mass ranges indicated.  Open squares show the median correction for each of the three samples in redshift bins.  It is clear that at the low redshift end of our sample (z$\sim$0.03) the aperture corrections can be large (a factor of $\sim$10) and they can be significantly larger for higher mass (larger) galaxies than for the typical galaxy.  By z$\sim$0.1, the median aperture correction has converged to a value of $\sim$3 and the median correction for higher mass galaxies is now not much larger than the median correction for galaxies of lower stellar masses, at the same redshift.

\begin{figure}
{\centering
\includegraphics[width=90mm]{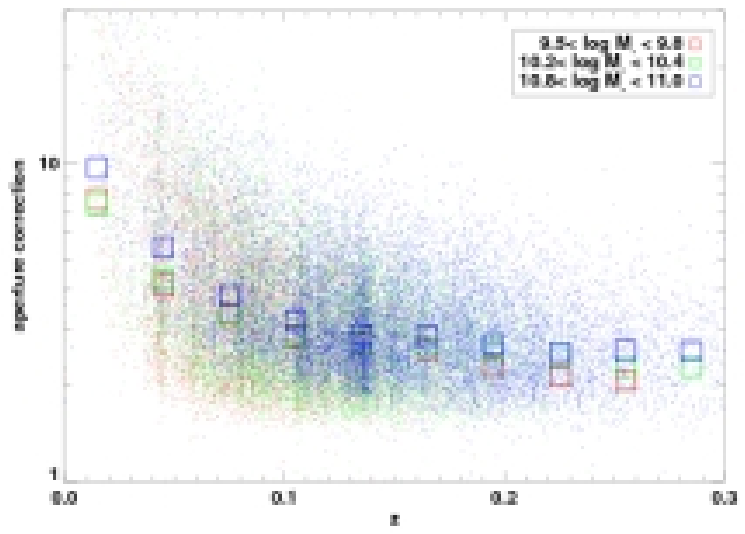}
\caption{The aperture correction, $u$-band Petrosian flux/fibre flux, as a function of redshift.  Coloured points show galaxies with stellar masses in the three mass ranges indicated.  Squares indicate the median corrections in mass bins for each of the samples.  
These show that although the aperture corrections for the most massive (and hence largest) galaxies 
are both large and  also even larger than the typical galaxy at lower redshifts (z$\lsim$0.1), towards the higher redshift half of our sample (0.1$\lsim$z$<0.2$) the corrections are typically a factor of three for all galaxies.  See text for discussion.
\label{fig:apcorr}
}
}
\end{figure}

\b04~derived an empirical colour-dependent aperture correction using $(g-r)$ and $(r-i)$ colours.  This empirical relationship is calibrated based on star-forming galaxies, but applied to galaxies of all spectroscopic classes.  \citet{Salim:2007rv} have cautioned that such a calibration does not seem to hold for non-star-forming classes, by comparing \b04~\sfrtot~with their GALEX-derived FUV SFRs (see in particular \citealt{Salim:2007rv}, fig.~5).  

We wish to test whether the difference seen at the high mass end (Fig.~\ref{fig:sfrd_lines}) between our emission line measures in Stripe 82 and \sfrtot~from \b04 (from a larger SDSS area) might be due to the aperture corrections used. To do this, first we directly compare our aperture corrections with \b04's aperture corrections (by dividing the \sfrtot~value in the MPA catalogue by the $SFR_{fibre}$ measurement to recover the correction used) on a galaxy-by-galaxy basis.  We plot the log of this difference as a function of stellar mass in Fig.~\ref{fig:cfapcorr}.  Larger filled points with error bars indicate the median and standard deviation of this measurement.  We see that, on average, our aperture corrections are systematically larger by $\sim$0.2dex than \b04, with significant scatter.  There is little trend with stellar mass, although the scatter becomes significantly larger at \ms$\gsim$10.0, with a large tail of objects towards lower differences (this work - \b04) and at \ms$\gsim$11.5 the median offset begins to increase.  This trend goes in the opposite direction to that required to explain the higher \b04 SFRD at high stellar masses versus ours, and so aperture corrections are unlikely to be a major effect.  

The fact that \b04~use a colour-dependent aperture correction and different measures of SFR (their $SFR_d$ and $SFR_e$) for galaxies of different spectral classes (but derived from star-forming galaxies) may lead to partially cancelling effects which result in us deriving comparable total SFRDs.  We reiterate that we use a single aperture correction for galaxies of all spectral classes and naively assume that all measured line luminosity is due to star-formation.  To check if this might be the case, we next compare SFRD estimates made entirely within the fibre (to remove the dependence on aperture corrections) and also examine these split by spectral type. 

Fig.~\ref{fig:sfrd_fibre} shows our best estimates of \ha~and \oii~SFRs in Stripe 82 with the aperture corrections set to unity (solid red and blue curves respectively). The black histogram shows \sfrtot~measured only within the fibre (which we refer to as \sfrfib.   The solid lines show these measurements just for galaxies classified as star-forming.  It is clear that \sfrfib~overestimates the SFRD with respect to our \ha~and \oii~estimates, even within the fibre.  For the star-forming class, \sfrfib~is most closely based on emission line luminosity (like the \ha~and \oii~methods) and so these should be expected to give the best agreement.  It can be seen that even for this class, the former indicator overestimates the latter by an amount comparable to the difference in aperture corrections (hence the total SFRDs give better agreement).  When all galaxy classes (rather than just star-forming) are considered, the agreement between the fibre-based measures is now somewhat better, at least for \ha~compared with \sfrfib.  It should be remembered that for non-star forming galaxies \sfrfib~does not directly use emission line luminosity, but rather a scaling based on SFR/stellar mass from the star-forming class.  Out best estimate of \oii~(using the empirical correction derived for star-forming galaxies) now underestimates the other two methods (as it does for total SFRDs in Fig.~\ref{fig:sfrd_lines}).  Again, this illustrates that contamination by AGN is unimportant for \oii-based measurements, and that the primary cause of the difference between \oii~and \ha-estimated SFRs is the sample selection described in \S\ref{sec:sampsel}.

Finally, an independent check of the validity of our aperture corrections is provided by the photometric SFR estimators.  The agreement between our photometric SFRD estimates and our \ha~SFRD at the high mass end reassures us that our aperture corrections are reasonable, but this is the regime where the effects of dust are most important and so a conspiracy between an underestimate of the extinction correction in the UV coupled with an underestimate of the aperture correction in the emission line SFRDs might artificially cause good agreement.  In higher redshift surveys (z$\sim$1, say), aperture corrections are relatively mild and it is only for this low redshift (fibre-based) comparison where the aperture corrections can reach factors of several for the largest galaxies that this is of concern.

\begin{figure}
{\centering
\includegraphics[width=90mm]{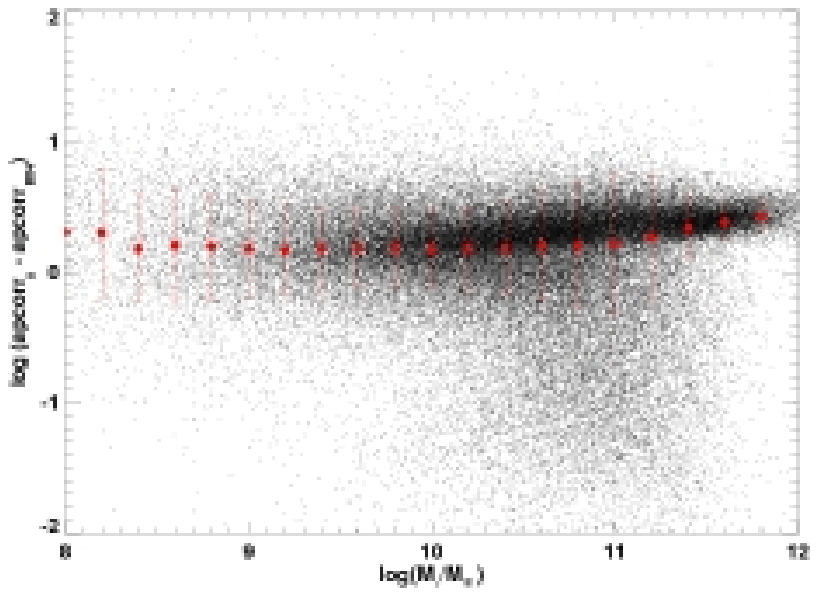}
\caption{Comparison of the $u$-band aperture correction with the colour dependent correction of \b04.  See text for discussion.
\label{fig:cfapcorr}
}
}
\end{figure}

\begin{figure}
{\centering
\includegraphics[width=90mm]{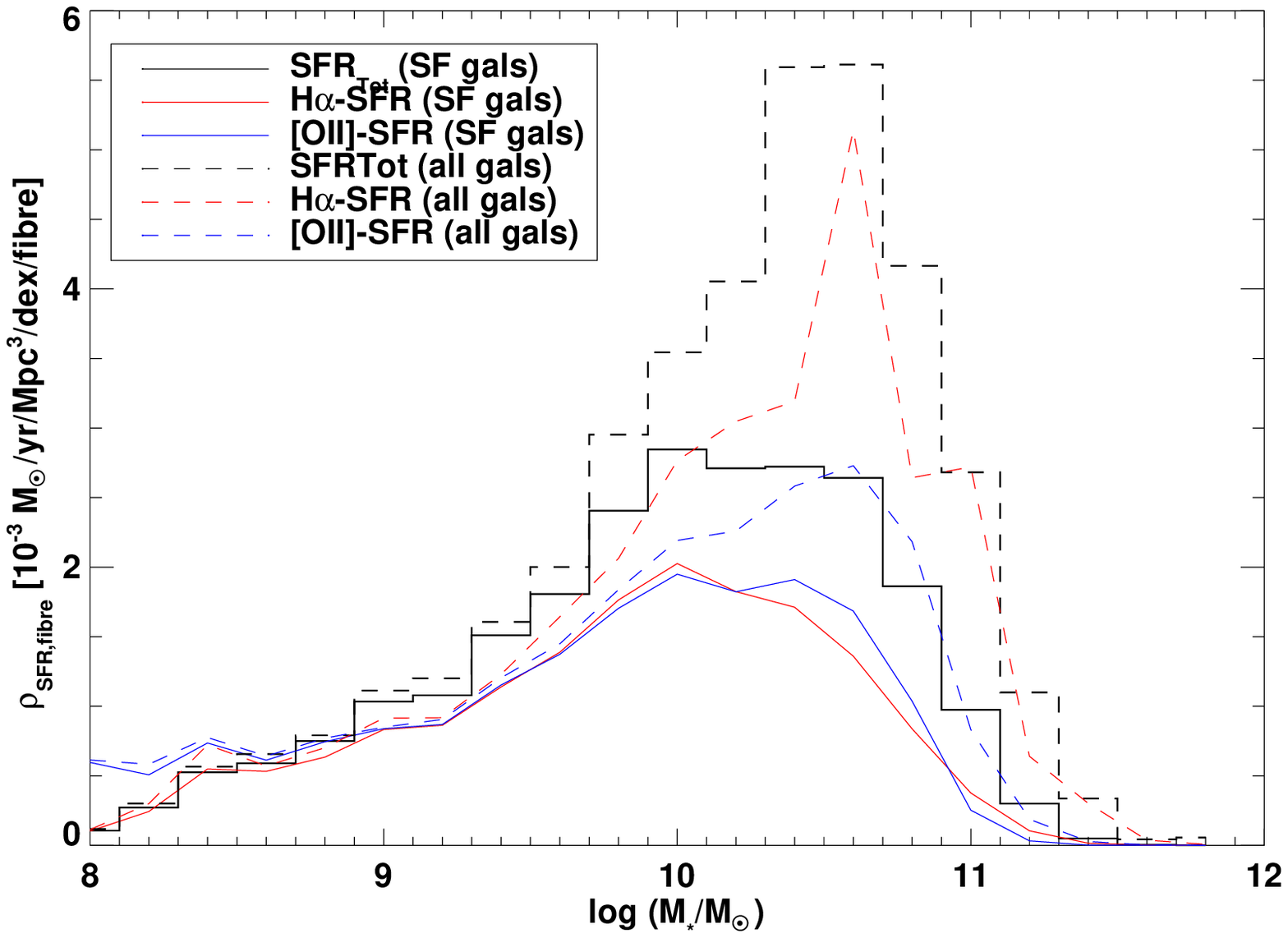}
\caption{SFRD as a function of stellar mass measured within the fibre for star-forming galaxies.  Red and blue curves show best estimates (Balmer decrement corrected and empirical correction) for \ha~and \oii~SFRs and black histogram shows that using \sfrtot. Only star-forming (\b04's class=1) galaxies are shown, to simplify the comparison between the indicators.  The SFR measurements within the fibre are in good agreement.  Dashed lines show the same indicators, but this time for galaxies of all spectral classes.  The \ha~and \oii-based indicators still predict slightly lower SFRs than \sfrtot~(which uses an alternate estimator for non-star forming galaxies).  See text for discussion.
\label{fig:sfrd_fibre}
}
}
\end{figure}

\section{completeness in stellar mass and evolution}
\label{sec:completeness}

At a low redshift limit of z$=$0.005 and a magnitude limit of $r$=17.77, the SDSS is complete in stellar mass to below \ms$\sim$8 (\b04).  For the current work, we impose a low redshift limit of z$=$0.032 in order to include \oii~in the observed spectrum.  The deeper Stripe 82 data impose a fainter magnitude limit of $r$=19.5. To estimate the completeness limit in stellar mass, we examine the stellar mass function in redshift slices.  The left panel of Fig.~\ref{fig:smass_z} shows the distribution of galaxies in Stripe 82 in stellar mass and redshift.  The dashed lines indicate the redshift limits we impose.  It is clear from the distribution of points that the lowest mass galaxies are only sampled at the lowest redshift range of the sample, due to the magnitude-limited nature of the sample.  In the right panel, the stellar mass function is constructed.  This follows the same Vmax procedure as the SFRD (Eqn.~\ref{eqn:vmax}), but replacing SFR with unity to simply count objects.  The log of the number density ($\log\phi$) is shown as a function of stellar mass (\ms) in $\Delta$z=0.01 redshift bins.  Error bars have been omitted for clarity.  It can be seen that the overall shape of the stellar mass function over the bulk of the mass range is very similar for most of the redshift range.  A notable exception is the very lowest redshift bin plotted, which shows a much lower $\log\phi$ at the high mass end.  This can also be seen as the paucity of high mass objects in the left panel, and may be due to cosmic variance, due to the relatively low volume sampled by the lowest redshift slice.  Nevertheless, the shape of the stellar mass functions are largely consistent (c.f., also \citealt{Bell:2003iq}) and so the position of the turnover at low masses may be used to estimate the completeness limit. The turnover in the lowest redshift bin seems to occur just above \ms$=$8.0 and so the survey should be complete to around this limit at this redshift.  This estimate may be somewhat pessimistic, as the stellar mass function considers galaxies of all spectral types/colours, whereas we primarily care about star-forming galaxies when estimating the SFRD.  Stellar mass incompleteness will occur first in a magnitude limited survey for non-star forming galaxies (which have redder colours and hence higher M/L ratios), so this should be considered as a lower limit to the completeness.  

\begin{figure*}
{\centering
\includegraphics[width=140mm]{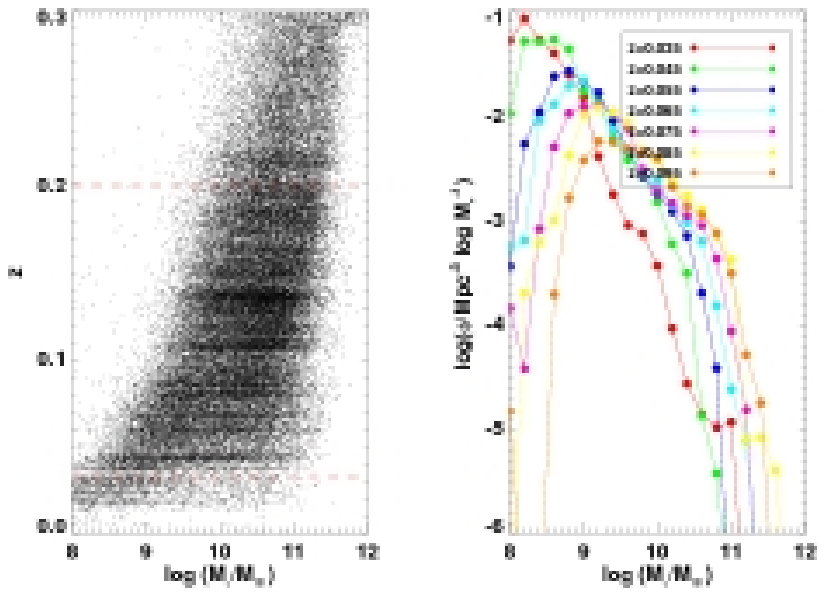}
\caption{Left panel: distribution of Stripe 82 galaxies in mass and redshift.  Dashed lines indicate redshift limits adopted for this survey.  Right panel: stellar mass functions in narrow redshift slices, in order to verify the limiting stellar mass probed by our 0.032$<$z$<$0.20 sample.  The turnover in the stellar mass function and the envelope of galaxies in mass vs redshift give an indication of the stellar mass completeness limit of the survey.  Only the redshift range z$<$0.1 is plotted, since the survey clearly becomes incomplete for the lowest mass galaxies by this point. See text for discussion.
\label{fig:smass_z}
}
}
\end{figure*}

The effect of evolution over the redshift range of the Stripe 82 data is of potential concern. No correction for evolution has been made to our SFRDs.  In order to assess this effect, Fig.~\ref{fig:sfrd_evol} shows the SFRD for \sfrtot, and the best corrected \ha~and \oii~measurements in two different redshift bins: $0.032<z<0.1$ (left panel) and $0.1<z<0.2$ (centre panel).  Even if the SFRs of individual galaxies are not evolving, the different redshift cuts imposed will result in different SFRDs, due to sampling effects, in the sense that the lowest mass galaxies must be drawn from the lowest redshift end of the survey (as they become too faint to be included in the magnitude selection of the survey at higher redshifts).  Also, since the volume probed over a fixed angular area survey will be larger at higher redshift, a larger fraction of the intrinsically rarer higher-mass galaxies might be expected to come from towards the higher redshift end of the sample.  The fact that our overall $0.032<z<0.20$ SFRD using \sfrtot is in good agreement with the evolution-corrected, $0.005<z<0.22$ SFRD of \b04~suggests that our whole sample comprises a fair representation of the local Universe 

In the higher redshift bin, the most obvious effect is the rapid decline of the SFRD at \ms$\lsim$9.5.  This is obviously due to incompleteness, since these lowest mass galaxies are only visible in the Stripe 82 data at the lowest redshifts in this survey.  At higher masses, the SFRD is comparable to that seen in Fig.~\ref{fig:sfrd_lines}, and the more rapid drop off than in the \b04~result is still present.  In the lower redshift bin, the low mass tail now agrees well with that of Fig.~\ref{fig:sfrd_lines}, but the high mass SFRD is suppressed.  This is likely due to evolution, in the sense that the universal SFR is declining towards the present day, and since the bulk of the SFRD occurs at higher stellar masses, the effect is most obvious there.  \b04~attempted to correct for evolution by assuming a universal power law evolution for the SFR of the form $(1+z)^3$ (see \b04~for the justification for this correction).  Thus, the  SFRD normalised at z$=$0.1, $SFRD_{0.1}$, can be calculated as $SFRD_{0.1} = SFRD_z (1+0.1)^3/(1+z)^3$, for a galaxy at redshift $z$ with an SFRD, $SFRD_z$.  The result of applying this correction is shown in the right panel of Fig.~\ref{fig:sfrd_evol}.  It can be seen that this is in good agreement with the results in Fig.~\ref{fig:sfrd_lines}, and so we conclude that evolution over the redshift range discussed here has little impact on our conclusions.  From the above discussion, assuming this form of a correction is approximately correct, the biggest systematic effect {\it not} applying this can have on our results is $\approx$10\% at the higher redshift end (and thus higher stellar masses), and $\approx$6\% at the lower redshift end (hence lower stellar masses).  This is comparable with the 6\% effect reported by \b04~(for a slightly wider redshift range).

\begin{figure*}
{\centering
\includegraphics[width=140mm]{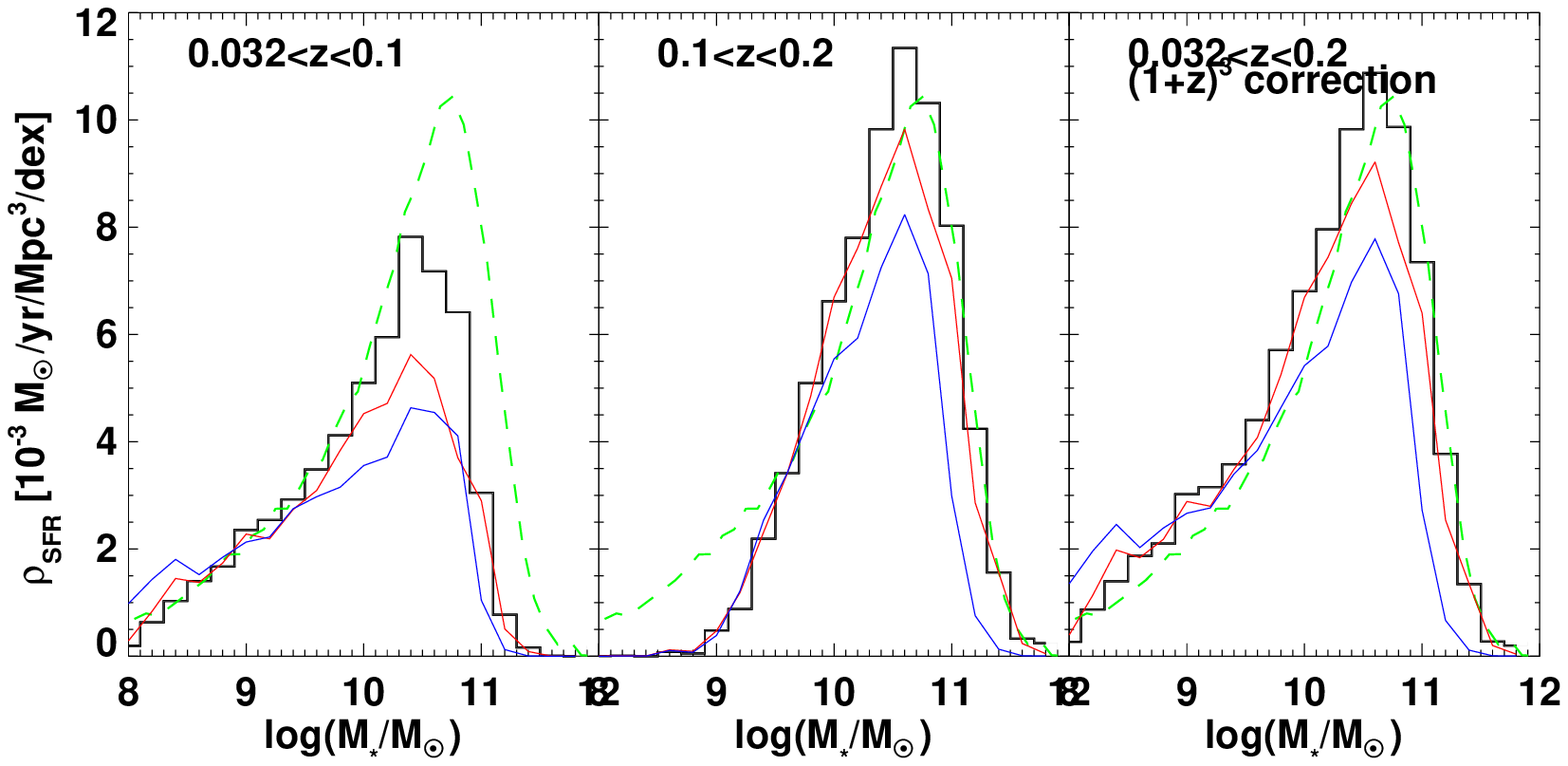}
\caption{The SFRD as a function of stellar mass, as for Fig.~\ref{fig:sfrd_lines}, but this time split into redshift bins to study the effect of evolution.  Red and blue curves show best estimates (Balmer decrement-corrected and best correction) for \ha~and \oii~SFRs respectively.  Solid histogram shows \sfrtot and green dashed surve shows \b04 SFRD determination (at z$=$0.1 in every case).  Left panel shows $0.032<z<0.1$, centre panel shows $0.1<z<0.2$ and right panel shows the full redshift range, but SFR of each galaxy has been corrected to z$=$0.1, assuming SFR scales as $(1+z)^3$.  See text for details.
\label{fig:sfrd_evol}
}
}
\end{figure*}

\label{lastpage}
\end{document}